\documentclass[]{emulateapj}
\usepackage{lscape}

\shorttitle{XTE J1814-338 burst oscillation variability}
\shortauthors{Watts, Strohmayer \& Markwardt}

\begin{document}

\title{Analysis of variability in the burst
  oscillations of the accreting millisecond pulsar XTE J1814-338}

\author{Anna L. Watts\altaffilmark{1}, Tod
  E. Strohmayer\altaffilmark{2} and Craig
  B. Markwardt\altaffilmark{3}} 
\affil{X-ray Astrophysics Laboratory, NASA Goddard Space
  Flight Center, Greenbelt, MD 20771, USA}

\altaffiltext{1}{National Research Council Resident Research
  Associate; anna@milkyway.gsfc.nasa.gov }
\altaffiltext{2}{stroh@clarence.gsfc.nasa.gov}
\altaffiltext{3}{Department of Astronomy, University of Maryland,
  College Park, MD 20772; craigm@milkyway.gsfc.nasa.gov}

\begin{abstract}
The accreting millisecond pulsar XTE J1814-338 exhibits oscillations
at the known spin frequency during Type I X-ray bursts.  The
properties of the burst oscillations reflect the nature of the thermal
asymmetry on the stellar surface. We present an analysis of the variability of the burst
oscillations of this source, focusing on three characteristics:
fractional amplitude, harmonic content and frequency.  Fractional
amplitude and harmonic content constrain the size, shape and position
of the emitting region, whilst variations in frequency indicate motion
of the emitting region on the neutron star surface.  We examine both
long-term variability over the course of the outburst, and short-term
variability during the bursts.  For most of the bursts, fractional
amplitude is consistent with that of the accretion pulsations,
implying a low degree of fuel spread.  There is however a population
of bursts whose fractional amplitudes are substantially lower,
implying a higher degree of fuel spread, 
possibly forced by the explosive burning front of a precursor burst.
For the first harmonic, substantial differences between the
burst and accretion pulsations suggest that hotspot geometry is not the only
mechanism giving rise to harmonic content in the latter. Fractional
amplitude variability during the bursts is low; we cannot rule out
the hypothesis that the fractional amplitude remains constant for
bursts that do not exhibit photospheric radius 
expansion (PRE). There are no significant variations in frequency in
any of the bursts except for
the one burst that exhibits PRE.  This burst exhibits a highly
significant but small ($\approx 0.1$Hz) drop in frequency in the burst
rise.  The timescale of the frequency shift is slower than simple
burning layer expansion models predict, suggesting that other
mechanisms may be at work.

\end{abstract}


\section{Introduction}

There are now seven known accreting millisecond pulsars:  SAX
J1808.4-3658 \citep{wij98, cha98}; XTE J1751-305 \citep{mar02}; XTE
J0929-314 \citep{gal02}; XTE J1807-294 \citep{mar03a}; XTE J1814-338
\citep{mar03b}; IGR J00291+5934 \citep{gal05}; and HETE J1900.1-2455
\citep{mor05}.  Of these systems only three, SAX J1808.4-3658
(hereafter J1808), XTE J1814-338 (hereafter J1814) and now HETE
J1900.1-2455 \citep{vand05}, have shown X-ray 
bursts.  The bursts of J1808 and J1814 show oscillations at or very close
to the
known spin frequency of the neutron star \citep{cha03,str03}.  This
suggests that rotation is modulating an asymmetry on the burning
surface that is near-stationary in the co-rotating frame.   The nature
of the asymmetry, however, remains unresolved. 

In the case of J1808 and J1814 we
know that fuel deposition is inhomogeneous, since otherwise we would not
observe the systems as pulsars.  One thing that is not clear, however, is whether the
material builds up at the deposition point or whether it spreads
out over the surface of the star,
thereby lessening asymmetries in fuel coverage \citep{ino99}.  In
either case, however, coverage is unlikely to be completely even.
This means that ignition is likely to occur at a particular point
rather than occurring simultaneously
across the entire surface.   As the burning
front spreads to engulf the available fuel (giving an overall rise in
luminosity), changes in the emitting region should be reflected in
the properties of the burst oscillations.   

There are several mechanisms that may give rise to a thermal
asymmetry.  If there are
significant inhomogeneities in fuel coverage, 
caused for example by magnetic channelling,
areas with more fuel should get hotter.  These thermal asymmetries
may then persist as the surface of the star cools in the burst tail.
If fuel coverage is even, the initial hotspot that develops at the
ignition point is expected to grow rapidly to engulf the whole star during the
burst rise.  If this is the case, what then causes the
asymmetry in the burst tail?  Possibilities include the development of
vortices driven by 
the Coriolis force \citep{spi01}, or a brightness pattern caused by
oscillations in the surface layers \citep{mcd87, lee04,
  hey04, lee05, pir05}.  Mechanisms that do not rely on fuel channelling are
of particular importance for the non-pulsing bursters, systems for
which there is  
 no evidence for asymmetric fuel deposition.  By attempting to
 distinguish the candidate mechanisms we hope to probe not only
 the nuclear burning process and the surface layer 
composition, but also the role of the magnetic field in controlling the
accretion flow and the motion of the surface layers.  

The transient system J1814 was first observed in outburst on June 3rd
 2003 by the {\it Rossi X-ray Timing Explorer} (RXTE) Galactic bulge
 monitoring campaign.  A longer 
 observation on June 5th confirmed that the source was a pulsar
 \citep{mar03b}.   The pulsar has a spin frequency of 314.36
 Hz, resides in a binary with an orbital period of 4.275 
 hr, and has a minimum companion mass of $\approx 0.15 M_\odot$
 (Markwardt et al 2005, in preparation).  J1814 is the
 widest and most massive binary of the known accreting millisecond
 pulsars: in this regard it is the most similar to the
 non-pulsing low mass X-ray binaries (LMXBs). The system remained in
 outburst until mid July 
 2003.  Over this period repeated observations were made
 with the RXTE Proportional Counter Array (PCA), during which 28 Type
 I X-ray bursts were 
 recorded.  A preliminary analysis of the bursts was given in 
 \citet{str03}.   

In this paper we present an
analysis of the variability of the burst oscillations of J1814.  We
focus on three particular characteristics:
fractional amplitude, 
harmonic content, and frequency.  Fractional amplitude and harmonic
content constrain the size, shape and position of the emitting region, whilst
any changes in frequency over and above those expected for orbital
corrections indicate motion of the emitting region on the stellar
surface.  We examine both short-term variations during
bursts, and long-term variations in burst properties over
the course of the outburst.  The short-term variability reveals how
the size, shape and position of the emitting regions evolve during the
thermonuclear burst.  The long-term variations reflect the
influence of the accretion and burning history.  This history
is likely to affect both the magnetic field (which may be suppressed as
the outburst proceeds, affecting fuel deposition; \citet{cum01}) and
the composition of the surface (as successive thermonuclear bursts
process the accreted material into heavier elements; \citet{taa80, woo04}).

Section \ref{var} gives an overview of our method of analysis.
In Section \ref{outburst} we review the general characteristics of the
outburst.  Sections \ref{obv} and \ref{ibv} detail our analysis of 
the variability of the burst oscillations.  Section \ref{obv} examines
variation in burst average properties over the course of the outburst,
whereas Section \ref{ibv} assesses variability during the bursts.  
Section \ref{disc} discusses the results and relates them to current
theories of burst oscillation mechanisms.  We conclude with brief
comments on issues requiring further study.

\section{Method of analysis}
\label{var}

\subsection{Computation of fractional amplitude and frequency}

Our analysis of J1814 was conducted using  125$\mu$s time
resolution PCA event mode data. Event mode data overruns, which are
often seen in the bursts of brighter sources, were not seen in any of
the J1814 bursts. We first barycentered the data  
using the JPL DE405 ephemeris and the source position determined from
PCA scans \citep{mar03b, kra05}.  Fractional amplitudes, frequencies and
harmonic content were then analysed using two complementary
techniques:  the  $Z_n^2$ statistic \citep{buc83, str02}; and pulse profile
fitting.  

The $Z_n^2$ statistic is very similar to the standard power spectrum
computed from a Fourier transform, but does not require that the event data
be binned.  It is defined as:

\begin{equation}
Z^2_n = \frac{2}{N} \sum_{k=1}^{n} \left[ \left(\sum_{j=1}^{N} \cos
  k\phi_j\right)^2 + \left(\sum_{j=1}^{N} \sin
  k\phi_j\right)^2 \right]
\label{Z}
\end{equation}
where $n$ is the number of harmonics, $N$ is the total number of
photons, and $j$ is an index applied to each photon.
The phase $\phi_j$ calculated for each photon is 

\begin{equation}
\phi_j = 2\pi\int_{t_0}^{t_j} \nu(t) dt
\label{phase}
\end{equation}
where $\nu(t)$ is the frequency model, and $t_j$ the arrival time
of the photon relative to some reference time.  The rms fractional
amplitude $r$ for the trial frequency is then given by

\begin{equation}
r = \left(\frac{\bar{Z}^2_n}{N_s}\right)^{1/2}
\label{r}
\end{equation}
where $N_s$ is the number of source
photons, and $\bar{Z}^2_n$ is the $Z_n^2$ statistic for the source
alone.  The statistic that we compute from the data, using equation
({\ref{Z}), includes all photons: $N_s$ from the source; and $N_b$ from
  the background.  If we assume that the background photons are not
  periodic for the range of trial frequencies investigated, the
  background makes no contribution to the term in the bracket in
  equation (\ref{Z}).  In this case 

\begin{equation}
\bar{Z}^2_n  = \frac{N}{N-N_b} Z^2_n
\end{equation}
Equation (\ref{r}) becomes

\begin{equation}
r = \left(\frac{Z^2_n}{N}\right)^{1/2} \left( \frac{N}{N-N_b}
\right)
\label{rcorr}
\end{equation}
The number of background photons, $N_b$, can be estimated using the
standard FTOOLS routine pcabackest and the PCA background models.   

We have now corrected for the background, but we have yet to
account for the effects of noise.  If the true
 signal power (as calculated using the $Z^2_n$ statistic) is
$Z_s$, then the measured values $Z_m$ will be distributed according to 

\begin{eqnarray}
p_n\left(Z_m: Z_s\right) & = & \frac{1}{2} \exp\left[-\frac{(Z_m + Z_s)}{2}\right]
\left(\frac{Z_m}{Z_s}\right)^{(n-1)/2} {}\nonumber \\ & & \times  I_{n-1} \left(\sqrt{ Z_m Z_s}\right)
\label{p1}
\end{eqnarray}
where the function $I_{n-1}$ is a modified Bessel function of the
first kind \citep{abr64}.  This
distribution can be derived following the procedure outlined by 
\citet{gro75} for binned data, modified to use the normalisation of noise power
used by \citet{lea83}, as discussed by \citet{vau94} and
\citet{mun02}.  Example distributions are shown in Figure
\ref{egdists}; note that the distribution differs significantly from a
normal distribution, particularly for weak signals.  We have confirmed
the accuracy of this distribution for a trial frequency model using
Monte Carlo simulations to generate events files where the model light
curve includes a periodic signal.  The expectation
value and variance of $Z_m$ are given by

\begin{equation}
\left< Z_m \right> = Z_s + 2n
\label{ev}
\end{equation}

\begin{equation}
\left< (Z_m - \left< Z_m \right>)^2 \right> = 4(Z_s + n)
\label{variance}
\end{equation}
The probability of obtaining a measured $Z_n^2$ that
lies between 0 and $Z_m$, given $Z_s$, is given by the associated
cumulative distribution function:

\begin{eqnarray}
f_n\left(Z_m: Z_s\right) & = &  1 -  \exp\left[-\frac{(Z_m +
  Z_s)}{2}\right]{}\nonumber \\ & & \times  \left[ \sum_{k=0}^{\infty}
  \sum_{l=0}^{k+n-1} \frac{(Z_s)^k (Z_m)^l}{l!k!2^{k+l}}\right]
\label{f1}
\end{eqnarray}
The probability of the true signal power lying between 0 and $Z_s$ given
a measured power $Z_m$ is then given by

\begin{equation}
f_n\left(Z_s: Z_m\right) = 1- f_n\left(Z_m: Z_s\right)
\label{f2}
\end{equation}
It is $Z_s$ that we want to use in equation (\ref{rcorr}); we use
equation (\ref{f2}) to infer this quantity from
$Z_m$.  In this paper we take the best estimate of $Z_s$ to be the value for which
$f_n\left(Z_s: 
Z_m\right)=0.5$.  This is a matter of choice, something that must be
  borne in mind when interpreting the results.  A reasonable
  alternative, for example, would be to pick the $Z_s$
  for which $p (Z_m:Z_s)$ is a maximum.  In computing the error bars
that appear in plots of 
fractional 
amplitude in this paper the errors on $Z_s$ are taken to be the points
$f_n\left(Z_s:
Z_m\right)=0.159$ and $f_n\left(Z_s:
Z_m\right)=0.841$ (equivalent to $\pm 1 \sigma$ for a normal
distribution).  

\begin{figure}
\centering
\includegraphics[width=8.5cm, clip]{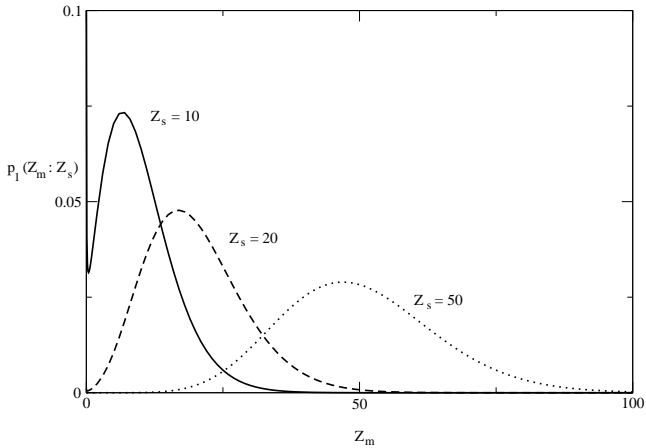}
\caption{Probability distribution $p_n\left(Z_m: Z_s\right)$ (equation
\ref{p1}), for 1 harmonic ($n=1$), showing the effect of varying
signal power
$Z_s$.  The distribution deviates significantly from a normal
distribution for low $Z_s$. } 
\label{egdists}
\end{figure}

One other issue that arises during the bursts is that there are two contributions
to the pulsations:  asymmetries due to accretion  (which give rise to the
pulsations observed between bursts) and asymmetries due to the
thermonuclear process. If 
we make the assumption that accretion continues during the burst,
the measured fractional amplitude will contain contributions from both
processes:

\begin{equation}
r = \frac{r_\mathrm{bur} N_\mathrm{bur} +
  r_\mathrm{acc} N_\mathrm{acc}}{N_s} 
\label{rcont}
\end{equation}
$N_\mathrm{bur}$ and $N_\mathrm{acc}$ are the number of
source photons arising from 
the burst and accretion processes respectively, with
$r_\mathrm{bur}$ and $r_\mathrm{acc}$ being the fractional
amplitudes of the two different processes.  The total number of source
photons, $N_s = N_\mathrm{bur} + N_\mathrm{acc}$.   If $N_\mathrm{bur}
\gg N_\mathrm{acc}$ then $r\approx r_\mathrm{bur}$, but we will not
always be in this regime, so we will need to estimate $N_\mathrm{acc}$
and $r_\mathrm{acc}$ in order to isolate $r_\mathrm{bur}$.  

An alternative method of analysis is pulse profile fitting. As with
the $Z^2_n$ statistic, we first pick a trial frequency model and assign a phase
to each event.  The events are then allocated
to phase bins; by recording the number of
events in each bin we can generate a pulse profile. To the resulting
binned data we fit a function of the form:

\begin{equation}
F(\phi) = A \left( 1 + \sum_{k=1}^n a_n \sin [k(\phi + b_n)]\right)
\end{equation}
where $n$ is the number of harmonics fitted.  The parameters $A$,
$a_n$ and $b_n$ are adjusted and the best fit 
found by minimising $\chi^2 = \sum_{i=1}^{N_\phi} [g_i -
  F(\phi_i)]^2/g_i$, where $\phi_i$ is the phase in the $i$th bin,
$N_\phi$ is the number of phase bins, and $g_i$ is the number of
events in that phase bin.  The rms fractional amplitude for that
trial frequency model for a specific harmonic is then given by
$a_n/\sqrt{2}$.  

All of the results presented in this paper were derived using the
$Z^2_n$ method.  However, we ran an extensive series of tests using
the profile fitting method, and verified that the results agreed to
within 1\% of those acquired using the $Z^2_n$ method. 

\section{General characteristics of the outburst}
\label{outburst}

In order to understand the variability of the X-ray bursts it is
important to understand the context in which they take place.   In
this section we review the general characteristics of the 2003
outburst insofar as they influence the bursts.

\subsection{Accretion rate}

A key factor affecting burst frequency and composition is the
local accretion rate (see \citet{str05} and
references therein). An indicator of variations in the accretion rate
is the non-burst flux, shown in the upper panel of Figure \ref{pers}.  \citet{gal04}
have used spectral fitting to estimate the bolometric 
flux for the 2003 outburst of J1814.  Using the distance estimate of
 $\approx 8$ kpc \citep{str03}, they infer that the peak accretion
during the outburst corresponds to  $\approx 4\%
\dot{M}_{\mathrm{Edd}}$.  A corresponding lower bound for the 
 local accretion rate, $\dot{m}$, can be calculated by assuming that
 material is deposited evenly across the stellar surface: $\dot{m} > 3
 \times 10^3 
 \mathrm{g~cm}^{-2}\mathrm{s}^{-1}$ (assuming a stellar radius of 10
 km and $\dot{M}_\mathrm{Edd} \approx 1.5\times 10^{-8} M_\odot
   \mathrm{yr}^{-1}$).  Accretion at this rate suggests that the
bursts of J1814 should be 
 mixed H/He bursts triggered by unstable helium ignition
 \citep{fuj81, fus87, cum00}.  For bursts taking place in the lower
 accretion rate regime after MJD 52830, one might expect both a drop in
 burst rate and more stable
 hydrogen burning, giving rise a higher helium fraction in the
 bursts. 

\begin{figure*}
\centering
\includegraphics[width=18cm, clip]{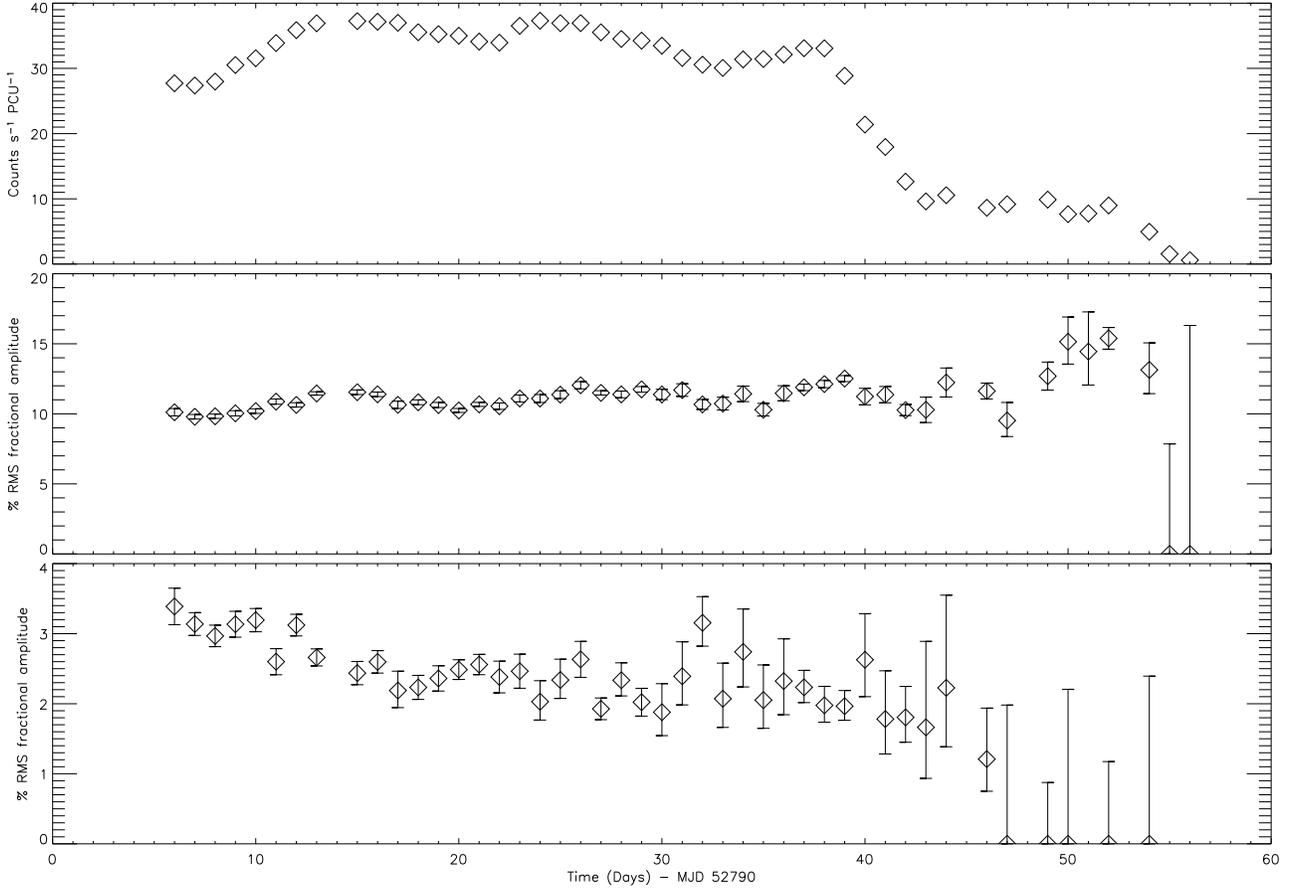}
\caption{Upper panel:  Daily average countrate in the energy band 2.5-25
    keV, corrected for background.
    Center panel: Daily average RMS fractional amplitude (\%) at the
    fundamental frequency of the pulsar.  Lower panel:  Daily average
    RMS fractional amplitude (\%) at the first harmonic of the pulsar
    frequency.}
\label{pers}
\end{figure*}

We note that
position in the color-color diagram is cited as a more reliable indicator
 of variations in accretion rate than X-ray flux \citep{van95}.
Analysis of the 2003 outburst of J1814 shows it to be in the
extreme island state, with the source moving to the left in a plot of hard 
vs soft color as the countrate declines after MJD 52830 \citep{van05}.
For brighter sources this would ordinarily signify an
increase in accretion rate, but for such a weak transient source movement
within the island state is not well studied, so we will assume
that the drop in flux represents a genuine 
drop in accretion rate.  

 Figure \ref{cov} shows the percentage daily coverage achieved by
 the RXTE PCA during the outburst, with the burst detection rate shown for
 comparison.  Given the coverage, the burst detection rate suggests a
 burst rate of $6-8$ per day in the peak of the outburst, with
 a lower rate of $4-6$ per day when the accretion rate is lower at the
 start and tail of the outburst (in accordance with the burst
 recurrence times suggested by \citet{gal04}).  A decrease in
 burst rate as accretion rate drops is in accordance with
 theoretical expectations, assuming that the fuel deposition footprint
 does not vary substantially over the course of the outburst.  

\begin{figure}
\centering
\includegraphics[width=8.5cm,clip]{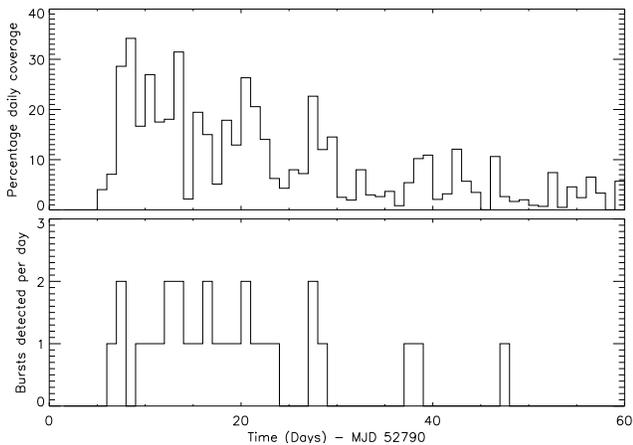}
\caption{The upper panel shows the percentage daily coverage achieved
    by the RXTE PCA during the 
    2003 outburst.  The lower panel shows the number of bursts
    detected per day.}
\label{cov}
\end{figure}

All of the bursts have rise times in
the range 1 to 8 s, with total burst duration 100 to 200 s.  This is
consistent with all of the bursts being
mixed H/He bursts.  However, the burst luminosity varies strongly from
burst to burst (Figure
\ref{burflux}).  The bursts with the shortest rise times
tend to be brighter (\citet{str03}; Figure \ref{risebur}), suggesting
that there may be
a higher proportion of helium in the burning mix.  A high proportion
of helium in the final burst is not unexpected (due to the lower
accretion rate), but it is notable that the brightness seems to vary even
during a regime of relatively constant accretion rate.  The burst
recurrence times are also variable \citep{gal04}.  Such variations could
occur naturally if different areas on the star are igniting.
Alternatively we could be seeing the effects 
of thermal or compositional inertia, where the burning history affects
subsequent bursts \citep{taa80, woo04}.  \citet{gal04} also noted that
burst recurrence times are shorter than would be
expected if accretion  were uniform across the
stellar surface.  This suggests that local accretion rate is higher
than the minimum value calculated elsewhere in this section, most
probably due to channelling of the accretion flow.   

\begin{figure}
\centering
\includegraphics[width=8.5cm,clip]{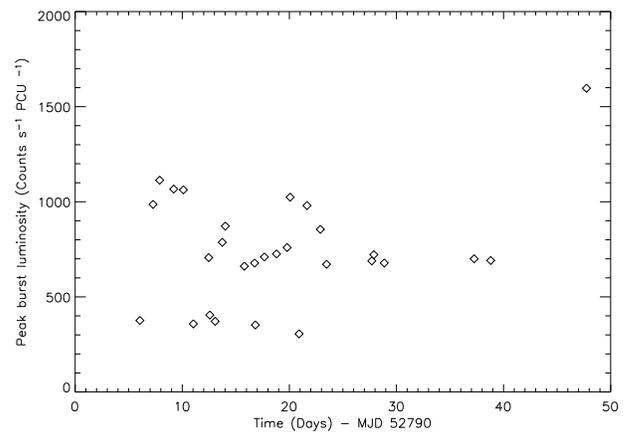}
\caption{The evolution of peak burst count rate (in the energy band 2.5-25
  keV, corrected for background and persistent emission) through the
  outburst.} 
\label{burflux}
\end{figure}

\begin{figure}
\centering
\includegraphics[width=8.5cm,clip]{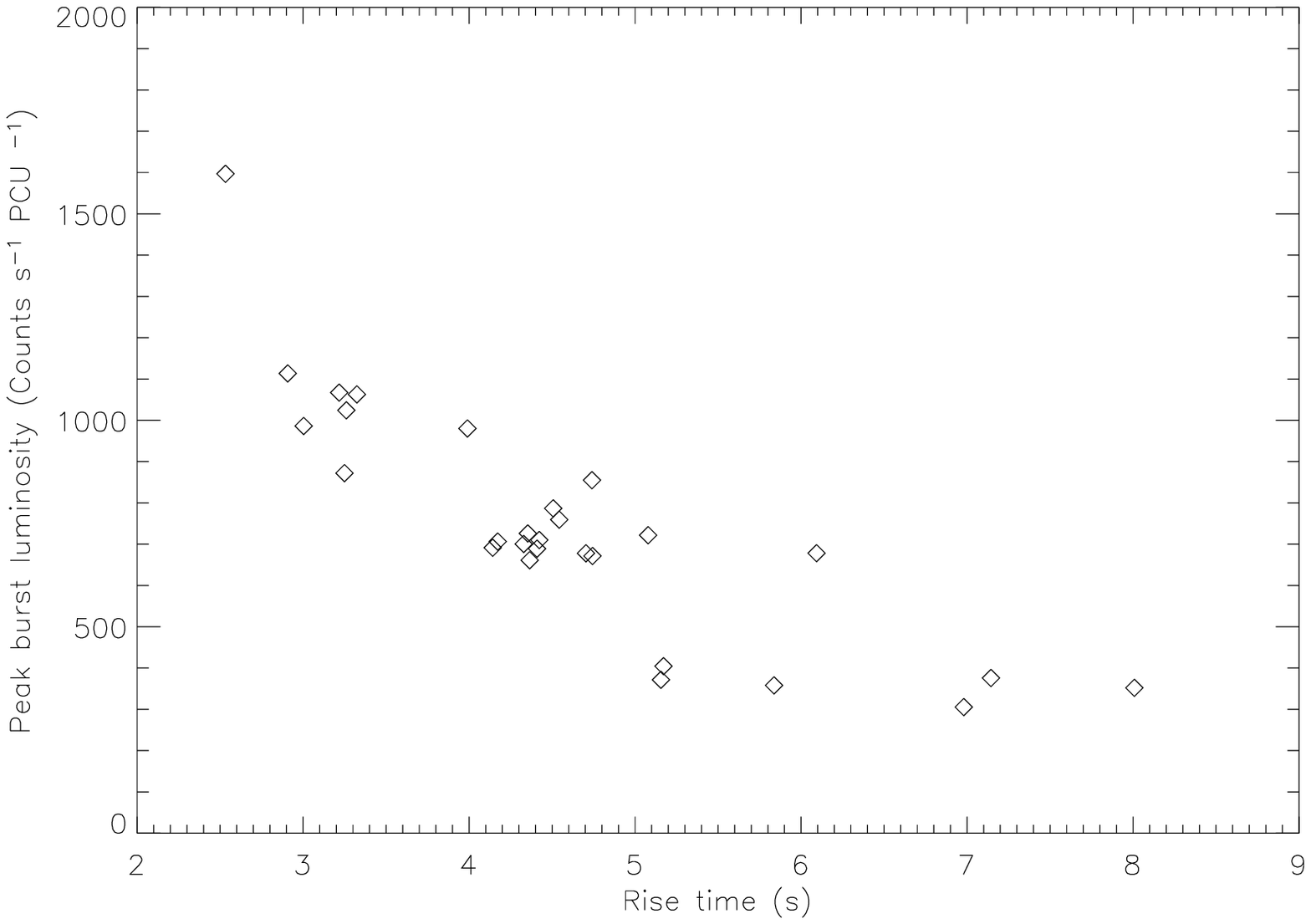}
\caption{Peak burst count rate (in the energy band 2.5-25
  keV, corrected for background and persistent emission) against burst
  rise time.  Rise time is defined as the time it
  takes the corrected count rate to rise from 10\% to 90\% of the peak
  value.} 
\label{risebur}
\end{figure}

\subsection{Fuel deposition pattern}

X-ray emission from an accreting neutron star contains contributions
from the accretion disk, the stellar surface, and the boundary layer
immediately above the surface.  Asymmetries in the resulting luminosity
pattern give rise to the non-burst pulsations.   To understand what
the fractional amplitude of these pulsations tells us about fuel deposition,
we need to understand the contribution of the different components to
the X-ray emission.  

In the 2003 outburst J1814 was in the island state, with relatively
low accretion rate. \citet{don03} argue that in this case the inner edge of the disk is likely to be far
from the star, in which case the bulk of
the emission from the accretion disk is likely to be outside the PCA
bandpass.  If the PCA emission is dominated by the contribution from
the stellar surface then the fractional amplitude and harmonic content
of the non-burst pulsations are a direct probe of the pattern of fuel deposition on
the stellar surface.  If on the other hand the disk contribution
cannot be neglected, the 
fractional amplitude of the surface component will be higher than that
measured (assuming that the disk emission is symmetric).  

Asymmetries are thought to arise because some of the matter from the
accretion disk 
is channelled out of the disk and along magnetic field lines towards
the magnetic poles.  The degree of channelling, which we define as the
difference between the minimum and maximum local accretion rates,  is a matter 
of debate.  Given the relatively weak magnetic fields of the accreting
millisecond pulsars, at least some of the matter is likely to penetrate the
field lines and accrete in a spherically symmetric fashion (see for
example \citet{sch78, gho79, spr90, mil98}).  X-ray generation will involve
thermal emission from the 
neutron star surface (as accreted matter is processed),
reflection of photons from the surface, and possible emission from
shocked plasma in the 
magnetic accretion funnel immediately above the stellar surface
\citep{bas76, lyu82, gie02, don03}.  Spectral analysis and 
modelling have been applied to J1808 in an effort to 
tease out the various contributions \citep{gil98, gie02, pou03}, but a
detailed study of this type has yet to be done for J1814.  

Regardless of the precise emission mechanism, however, the
fractional amplitude of the non-burst pulsations
will reflect the amount of material deposited in the region of the
magnetic polar
caps as compared to the rest of the stellar surface.    The effect of
different deposition geometries is illustrated in Figure
\ref{geometry}.  If the
fractional amplitude is very high we must have a bright hotspot on an
otherwise dim surface (strong channelling), and the geometry must be such
that the hotspot moves out of the field of view as the pulsar
rotates (upper panel, Figure \ref{geometry}).  If fractional amplitude
is lower we have two 
possibilities.  The first is that channelling is weak, leading to 
reasonably bright emission from much of the stellar surface.  In this
case there will be no major changes in brightness as the pulsar rotates (center panel, Figure \ref{geometry}).  The second is that
channelling is strong and most of 
the star is dim, but
that spot size and observer inclination are such that the projected area
of the hotspot is never zero (lower panel, Figure \ref{geometry}).

\begin{figure}
\centering
\includegraphics[width=8cm, clip]{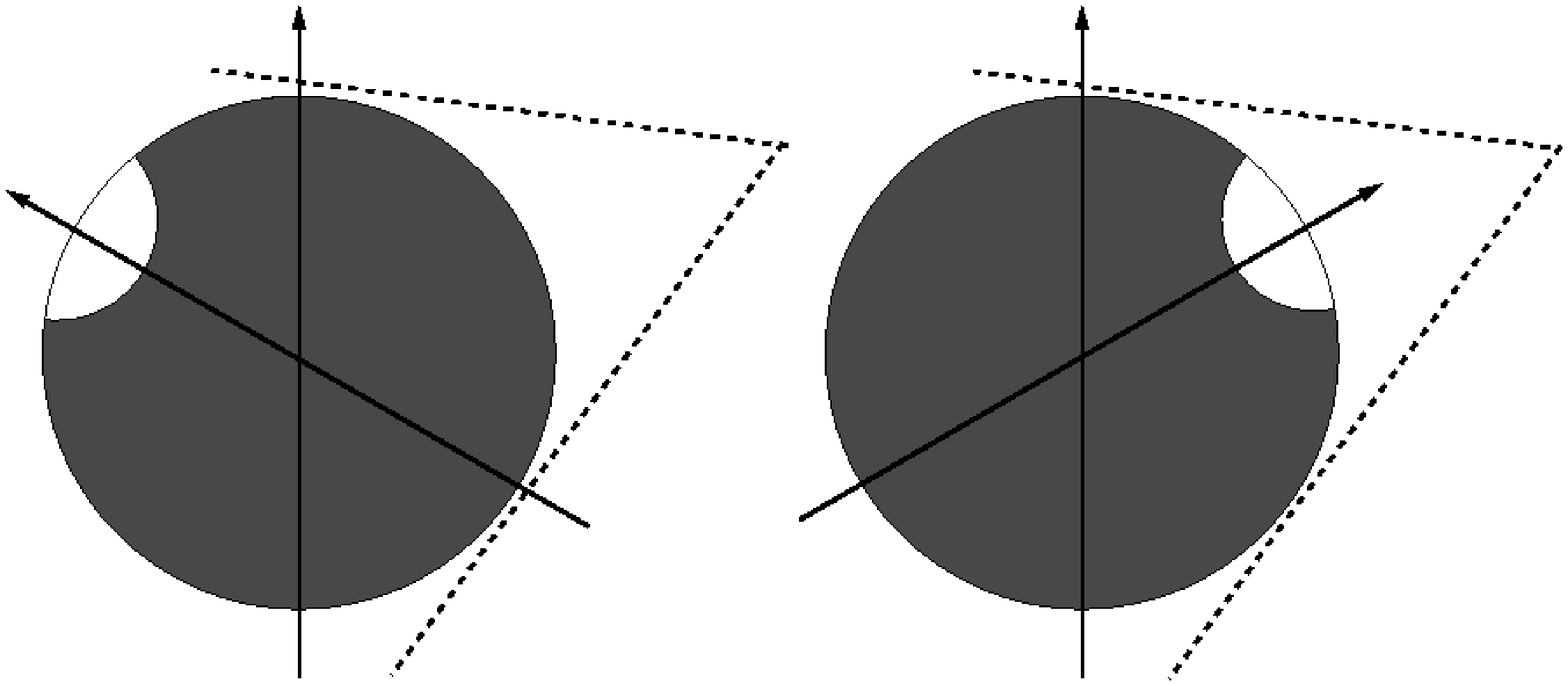}
\includegraphics[width=8cm,clip]{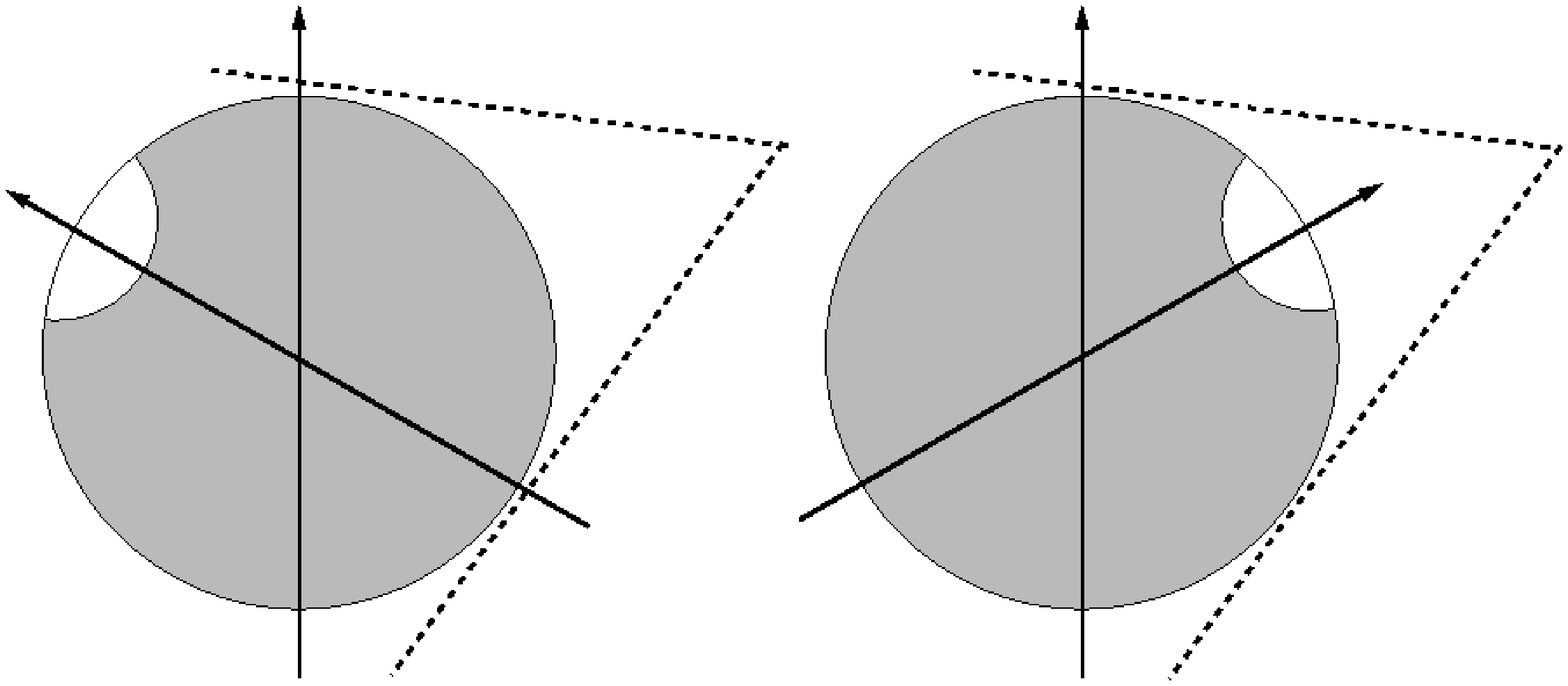}
\includegraphics[width=8cm,clip]{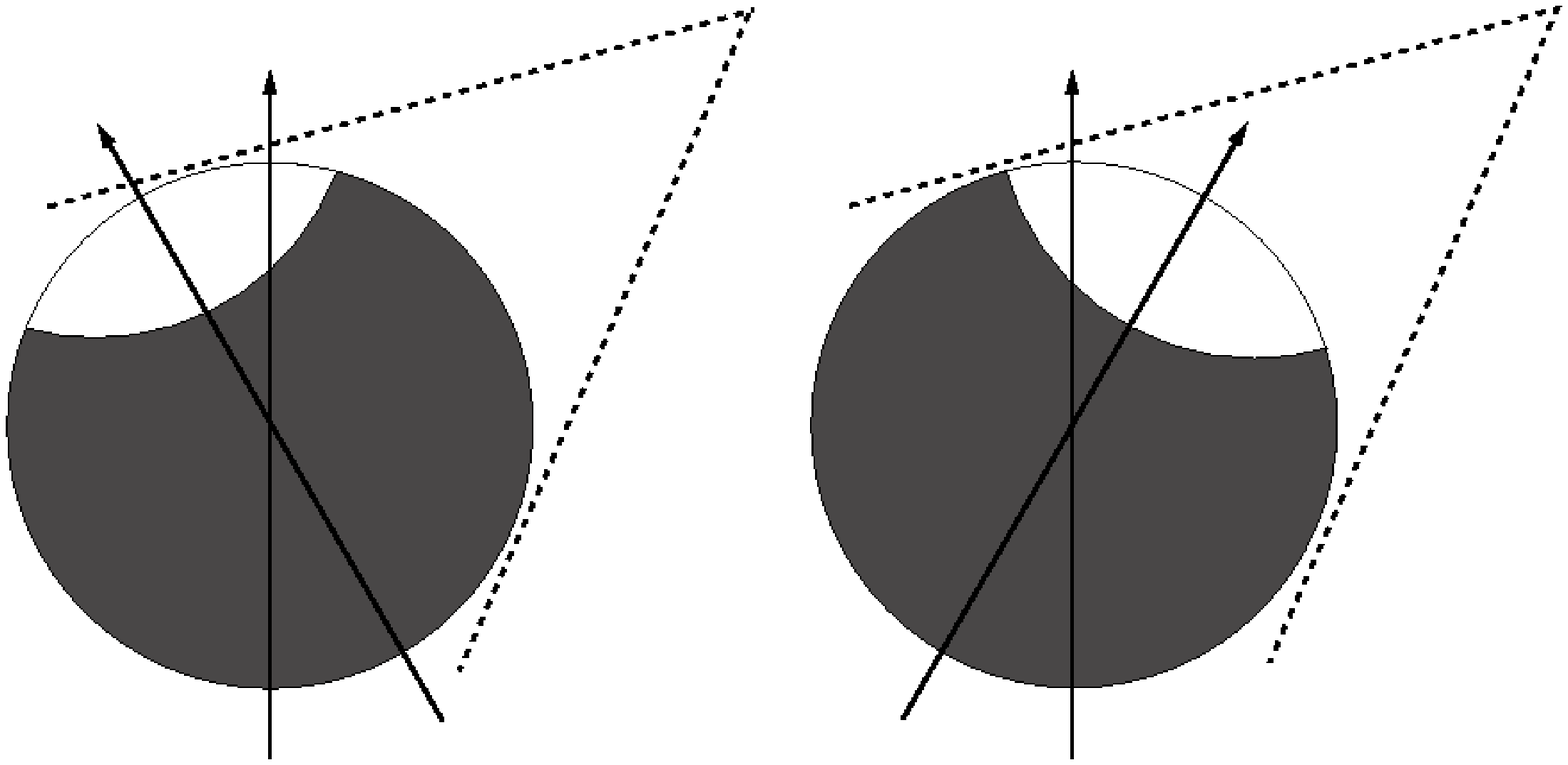}
\caption{The effects of different geometries and degrees of magnetic
  channeling on non-burst pulsation amplitude.  The
  dashed lines represent the field of view of 
  the observer.  The vertical arrow marks the rotation axis, the
  inclined arrow the magnetic dipole axis.  The white area is the
  visible magnetic polar cap (the antipodal cap is not shown).  The
  rest of the star is shaded either dark or light grey depending on
  its luminosity.  The left and right panes in each of the three
  panels indicate how the observer's view
  changes as the 
  pulsar rotates.  In the upper panel, magnetic channeling is strong.
  The bulk of the material falls onto the magnetic polar caps, giving
  a high luminosity in this region.  The rest of the star has very low
  luminosity.  The geometry is such that the hot magnetic cap rotates
  out of the field of view, giving a high fractional amplitude.  In
  the center panel, the geometry remains the same, with the spot
  rotating out of the field of view.  This time however, magnetic
  channeling is weaker and the rest of the star is correspondingly
  brighter.  Fractional amplitude will be lower.  In the lower panel
  magnetic channeling is again strong, but this time the geometry
  (spot size and observer inclination) is such that some
  portion of the hot cap is always in view. Fractional amplitude will
  again be lower.}
\label{geometry}
\end{figure}

The strength of any harmonics may allow us to distinguish these
possibilities.  Harmonic content may arise in several ways:  firstly if
the geometry is such that the second antipodal magnetic cap is also
visible; secondly due to the Doppler shifts,  which are more
pronounced if the spot is close to the 
rotational 
equator;  and thirdly due to the effects of beaming or a non-spherical
accretion footprint. 

The lower two panels of Figure \ref{pers} show the evolution of
fractional amplitudes of the fundamental and first harmonic of the
pulsar frequency during the non-burst emission.   These values were
calculated according to the prescription outlined in Section
\ref{var} (see \citet{cui98} for similar plots for J1808).  Daily
values of $Z_m$ were calculated using all good events 
from a given day.  The frequency 
model used was the best fit binary 
orbital ephemeris (Markwardt et al 2005, in preparation). The
fractional amplitude of the fundamental, at $\approx 10$ \%, is
relatively low, suggesting one of the two possible low amplitude
geometries.   

A comprehensive study of variability in the accretion pulsations is beyond the
scope of this paper. The analysis techniques used
could certainly be applied to the accretion pulsations, however, and
such an investigation could be a valuable 
diagnostic of the accretion flow.  The results related to the
accretion pulsations 
that we do present are included for two reasons. Firstly, we make the
assumption that   
the accretion process that gives rise to the non-burst pulsations
continues even during the bursts.  We need to take into account this
contribution to the overall fractional amplitude when calculating the
fractional amplitudes due to the thermonuclear burst process.
Secondly we wish to compare the fractional amplitudes and harmonic
content of the bursts with those of the non-burst pulsations.  If
fractional amplitude is substantially lower during the bursts, for
example, this would suggest that fuel is spreading over the surface of
the star after accretion.

\section{Variability over the course of the outburst}
\label{obv}

In this section we consider variation in the average properties of each
burst over the course of the outburst.  We define the burst start and
end times as being the times between which the count rate consistently
exceeds the average persistent rate.  All photons arriving between
those times are considered to be associated with the burst, and the
analysis methods are applied to the set of photons as a whole.  Table
\ref{bdatatab} summarises average burst properties for the set of
bursts.  

\citet{str03} showed that the burst oscillation frequency was
consistent with the pulsar frequency, and that frequency shifts during
the bursts were very small.  In calculating burst average fractional
amplitudes we therefore neglect any additional frequency shifts and
take as our frequency model the best fit orbital ephemeris.  We will
discuss frequency variability during the bursts in more detail in
Section \ref{ibv2}.  

 To isolate the fractional amplitude associated with the burst
 process, one needs to know the accretion luminosity and the accretion
 fractional amplitude (equation \ref{rcont}).  We can estimate these
 quantities by analysing segments of data immediately 
 preceding or following the bursts. One problem with this 
 approach is that one has to assume that the accretion flux and
 fractional amplitude remain steady over timescales of $\sim 100$~s.
 Analysis of  
 the non-burst pulsations indicates that statistical fluctuations
 alone can lead to these quantities varying routinely by 10-20\%. For
the fainter bursts, where $N_\mathrm{acc} \approx N_\mathrm{bur}$,
 this could introduce errors of this magnitude into our calculation of
 burst fractional amplitude.

To minimise the effect of variations (statistical or intrinsic) in
the accretion related pulsations we calculate fractional
amplitudes using only events that occur when the flux is at least
twice the pre-burst average.  By 
setting such a threshold we ensure 
that $N_\mathrm{acc}/N_\mathrm{bur} < 0.25$ for all of the bursts in
our sample (see Table \ref{bdatatab}).  This makes $r$ a better 
approximation to $r_\mathrm{bur}$, and reduces the effect of
variations in $N_\mathrm{acc}$ and $r_\mathrm{acc}$.  Accretion
corrections will still be important, however, if the estimated $r_\mathrm{acc}$
differs substantially from the measured $r$.
  
Figure \ref{fun} shows the burst fractional amplitude at the
fundamental frequency of the oscillations over the course of the
outburst.  Corresponding results for the
first harmonic are shown in Figure \ref{fir}. Accretion corrections 
have little effect on the fundamental fractional amplitudes, but they
have a noticeable effect on the first harmonic fractional amplitudes.

\begin{figure}
\centering
\includegraphics[width=8.5cm,clip]{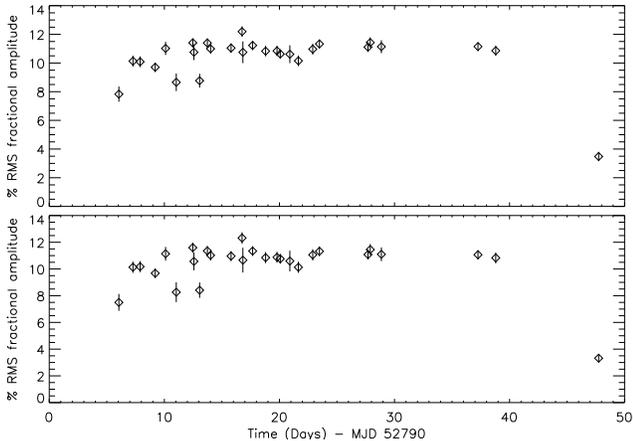}
\caption{Burst average fractional amplitude at the fundamental frequency of
  the  oscillations, showing variation over the course of the
  outburst.  The fractional amplitude is calculated
  using only events arriving between points when the flux is at least
  twice the pre-burst level.  The values in the upper panel are not
  corrected for accretion; the lower values are corrected.  The effect
  of including the accretion correction is for the most part imperceptible;
  the largest changes (a reduction of 4-5\%
  of the uncorrected value) are to Bursts 1, 6, 9 and 28.}  
\label{fun}
\end{figure}

\begin{figure}
\centering
\includegraphics[width=8.5cm,clip]{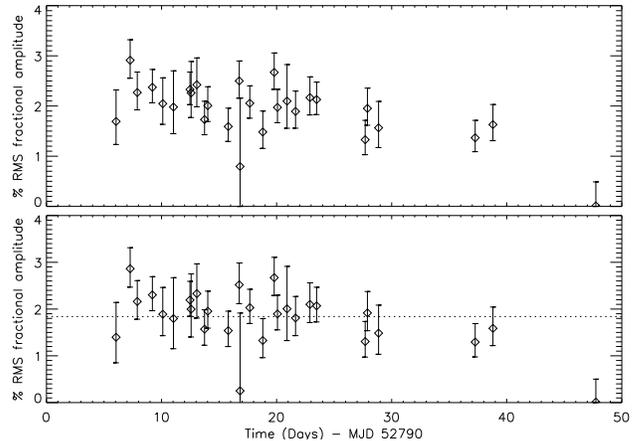}
\caption{Burst average fractional amplitude at the first harmonic of
  the  oscillations, showing variation over the course of the
  outburst. The fractional amplitude is calculated
  using only events arriving between points when the flux is at least
  twice the pre-burst level.  The values in the upper panel are not
  corrected for accretion; the lower values are corrected.  In this
  case the effect of including the accretion contribution is
  noticeable.  For ten bursts (1, 5, 6, 7, 8, 10, 14, 16, 25 and 26)
  the fractional amplitude drops by more than 5\% of the uncorrected
  value.  The dashed line in the lower panel shows the best fit
  constant fractional amplitude for the first 27 bursts. } 
\label{fir}
\end{figure}

\subsection{Testing the hypothesis that fractional amplitude remains constant}
\label{obv1} 

Let us first consider whether there is genuine variation in the
burst fractional amplitude over the course of the outburst. In other
words, are the data consistent with a burst fractional amplitude that
remains constant over the course of the outburst?  We start by
defining a simple measure of amplitude variability: 

\begin{equation}
\bar{\chi}^2 = \sum_i^{N_\mathrm{bur}} (r_i - r_\mathrm{model})^2
\label{barchi}
\end{equation}
$N_\mathrm{bur}$ is the number of bursts, $r_i$ is 
the fractional amplitude calculated for a particular burst, and
$r_\mathrm{model}$ is the trial amplitude model, in this case a
constant. We compute this quantity for the data for a range of trial fractional
amplitudes.  In
contrast with the more familiar $\chi^2$ statistic we do not
weight by the errors because they are non-Gaussian and non-symmetric.
Instead we will use Monte Carlo simulations to determine the distribution of
$\bar{\chi}^2$ under the constant fractional amplitude hypothesis and estimate the
significance of the result.  

We therefore need to simulate $N_\mathrm{sim}$ sets of $N_\mathrm{bur}$
fractional amplitude measurements.  For each burst $Z_s$, and hence
the distribution of fractional amplitudes, is different (because $N_s$
and $N_b$ are different), the distribution being wider for lower
$Z_s$.  Rather than generating artificial events files for each burst,
we can make  use of the fact that we know the
cumulative distribution function $f(Z_m:Z_s)$.  If we are running
$N_\mathrm{sim}$ simulations we start by generating $N_\mathrm{bur} \times
N_\mathrm{sim}$ random numbers drawn from a uniform distribution
between 0 and 1.  These are our values of $f$.  For each burst we take
$N_\mathrm{sim}$ of these $f$ values, compute the appropriate $Z_s$ for the
trial fractional amplitude, and solve equation (\ref{f1}) numerically
to generate $N_\mathrm{sim}$ values of $Z_m$. We then proceed exactly
as we did for the data and compute 
fractional amplitudes.  We have now simulated the distribution of $r$
for each burst under the constant fractional amplitude hypothesis.  

We then group the simulated data into sets of $N_\mathrm{bur}$,
each set containing one point from each different burst, and compute
 $N_\mathrm{sim}$ values of $\bar{\chi}^2$.  We repeat
this exercise for the full range of trial fractional amplitudes.  
By computing the number of simulations in which the simulated
$\bar{\chi}^2$ exceeds the measured value we can determine the
significance of the measured variability.

We start with the fractional amplitude at the fundamental frequency of
the oscillations.  If we consider the full set of 28 bursts we find
that the simulated 
$\bar{\chi}^2$  does not exceed
the measured value once even after 
simulating $10^4$ sets of bursts. This result holds true for both
accretion corrected and non-corrected values, and constitutes a strong rejection
of the constant fractional amplitude hypothesis.

What about relaxing the hypothesis?  The most obvious outlier is the
data point from the last burst.  This burst takes place when the  
accretion rate has dropped substantially, and it is plausible that
the fractional amplitude may indeed change for this burst.
\citet{str03} also found evidence that this burst exhibits
photospheric radius expansion, which would have the effect of
suppressing the fractional amplitude.   We therefore relax our
hypothesis and ask how likely it is that all of the bursts 
except for the last one have a constant fractional amplitude. This
reduces the measured  $\bar{\chi}^2$ substantially, but the simulated 
 $\bar{\chi}^2$ still never exceeds the measured
value once even after $10^4$ simulations. The constant fractional
amplitude hypothesis is still rejected.  There is genuine variation in
burst fractional amplitude.  

Note that in this analysis we have taken no account of the 
temporal order of the measurements, despite the fact that there seem
to be more values below the best fit fractional amplitude at the start
of the outburst. We are reluctant to make any pronouncement about
whether or not this is statistically 
significant.  Our caution stems from the fact that RXTE is likely
to have missed many bursts, particularly in the second part of the
outburst, where daily coverage dropped dramatically (see Figure
\ref{cov}). To make any progress in a temporal analysis one would need
even coverage across the outburst.  

For the first harmonic the simulated $\bar{\chi}^2$ exceeds the
measured value for the full set of bursts in 31 simulations out of
1000 (for the non-corrected values), the best fit amplitude being
1.95\%.  This constitutes a rejection of 
the hypothesis at a level just 
short of  2$\sigma$.  If we exclude the last burst from the set, 
the simulated $\bar{\chi}^2$ exceeds the measured value in 390
simulations out of 1000
simulations, with best fit amplitude 2\%, and the hypothesis cannot be
ruled out at all.  The 
strength of the rejection in both cases falls if we include the
accretion correction:  the best fit amplitude excluding the last burst
is shown in Figure \ref{fir}.  

We make one additional comment.  In
the later part of the outburst there are fewer bursts above the best
fit line than in the earlier stages.  \citet{bha05} noted this
apparent reduction in harmonic content in their paper.  One should
however be very cautious here because of the fall in coverage in the
later phases of the outburst.  Excluding the last burst (which takes
place in a dramatically different regime) the data are well fit by a
constant fractional amplitude, and confirmation that there is a
definite drop in harmonic content requires more observations in the
tail of the outburst.  

\subsection{Testing the hypothesis that burst fractional amplitude
  matches accretion fractional amplitude}
\label{obv2}

We also want to test whether the burst fractional
amplitudes are the same as the daily
average accretion fractional amplitude at the time of each
burst.  In
other words, we want to test the hypothesis
that the fractional amplitude remains constant for a given day
irrespective of the emission 
mechanism.  This would indicate a low degree of fuel spread from the
deposition point.

In testing this hypothesis we must take into account the fact that 
there is uncertainty in our measurement of daily average fractional
amplitude, as well as in our measurements of burst fractional
amplitudes.  We proceed as follows. For a given day we have one or
two values of fractional amplitude calculated during bursts, and one
value of fractional amplitude
formed by folding together non-burst data.  We start by
identifying, for each day, the fractional amplitude $r_m$ that maximises the
likelihood of obtaining both the burst and non-burst measurements made 
during that day.  The likelihood $L$ is defined as: 

\begin{equation}
L = \prod_i  p_i(Z_m: Z_s(r, N, N_b))
\end{equation}
where the product contains the two or three data points that we have for each
day \citep{wal03}.  

Having identified the most probable fractional amplitude $r_m$ for each
day, we proceed as we did for the constant fractional amplitude case,
computing a variability measure for the data and comparing it to
simulations.  This time however we must include the non-burst data
points in our test:   both the burst and non-burst measurements
are equally valid tests of the hypothesis that daily fractional
amplitude is constant irrespective of mechanism.

We compute two different measures of variability.  The first is just
 an amended version of the 
 $\bar{\chi}^2$ defined in equation (\ref{barchi}):

\begin{equation}
\bar{\chi}^2 = \sum_i^{N_\mathrm{bur}+N_\mathrm{day}} (r_i -
r_\mathrm{model})^2 
\end{equation}
where $N_\mathrm{day}$ is the number of days on which bursts were
observed, for which we
have computed non-burst fractional amplitude measurements.  In this case
$r_\mathrm{model}$ is the value of $r_m$ appropriate to each data
point.  This measure checks overall scatter of burst and
non-burst points.  It does not take into account the sign of the
difference.  As such, it will not pick out unusual behavior such as the
burst amplitudes 
always lying below the best fit value, with the accretion values
always lying above. For this reason we also compute a second
measure for the burst values alone:  

\begin{equation}
\bar{\chi} = \sum_i^{N_\mathrm{bur}} (r_i - r_\mathrm{model})
\label{barchione}
\end{equation}
As before, we run Monte Carlo simulations to determine the
distribution of both of our measures under the assumption that the
hypothesis is correct. 

Figure \ref{pbfunag} shows the degree of overlap between the burst and
non-burst fractional amplitudes for the fundamental.  Performing the
analysis on the full 
set of bursts, the hypothesis that the bursts have the same fractional
amplitude as the persistent emission is rejected at a level greater
than one part in $10^4$. The figure illustrates why this is the case.
For most of the bursts, 
there is in fact rather good agreement between the burst and non-burst
fractional amplitude.  There are however 6 bursts with fractional
amplitudes that are substantially lower than the persistent fractional
amplitude.  These are bursts 1, 6, 9, to a lesser extent bursts 26 and
27, and of course burst 28. 

\begin{figure}
\centering
\includegraphics[clip, angle=270, width=9cm]{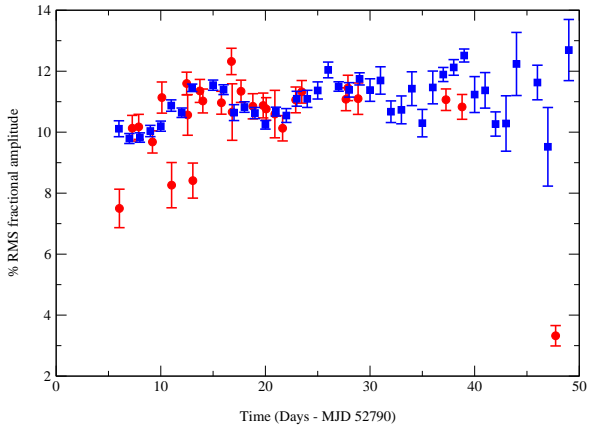}
\caption{The squares (blue) mark the non-burst daily average fractional
  amplitude of the fundamental; the
  circles (red) the burst average fractional 
  amplitude.  For most of the bursts there is relatively good
  agreement between burst and non-burst values, but for Bursts 1, 6,
  9, 26, 27 and 28 the fractional amplitude of the bursts is far lower
  than that of the non-burst pulsations. } 
\label{pbfunag}
\end{figure}

\begin{figure}
\centering
\includegraphics[clip, angle=270, width=9cm]{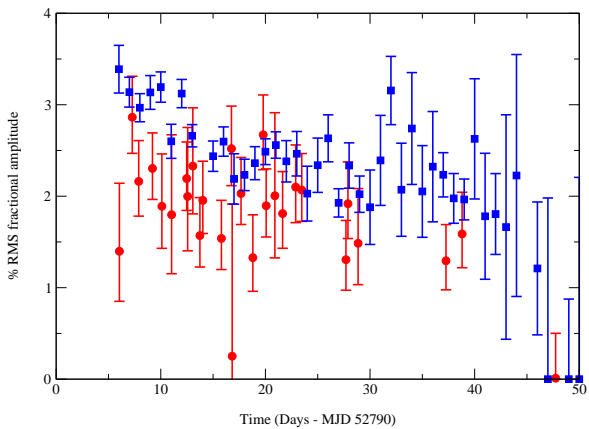}
\caption{The squares (blue) mark the non-burst daily average fractional
  amplitude of the first harmonic, the circles (red) the burst average
  fractional 
  amplitude.  The accretion-corrected burst fractional amplitudes are
  consistently lower 
  than the accretion values.  } 
\label{pbfirag}
\end{figure}

The fractional amplitude in burst 28 is
thought to be suppressed because of photospheric radius expansion.  If
we assume that the true fractional amplitude at the surface is simply
obscured by the photosphere, we can re-compute fractional amplitudes
neglecting events from the peak of the burst. In this case we obtain a
burst average fractional amplitude that is higher (at approximately
6\%), but which is still substantially lower than the persistent value.
A simple explanation for these bursts is that the fuel has spread
further from the deposition point.  For Burst 28, where recurrence
time has increased, this is perhaps reasonable.  For the other bursts,
it is hard to understand why only a subset of the bursts are affected
if the only factor affecting spread is the time elapsed since the last
burst.  One  possibility is that the
explosive burning front of the previous burst has forced fuel out
across the stellar surface.  The nature of the precursor burst would
then be an important determining factor, but this is hard to verify
given the gaps in the current data.  

Figure \ref{pbfirag} shows the degree of overlap between burst and
non-burst fractional amplitudes for the first harmonic.  Our analysis
shows that once again the hypothesis that the burst have the same
fractional amplitude as the persistent emission can be rejected at a
level greater than one part in $10^4$.  The most stringent rejection
is provided by the 
$\bar{\chi}$ test (equation \ref{barchione}).  It is clear from the
figure that the fractional
amplitude of the first harmonic in the bursts is, for all but 3 of the
bursts, lower than the non-burst amplitude.  Moreover, the trend of 
reducing first harmonic content seen in the non-burst pulsations is
not reflected in the burst oscillations.  This suggests that some
mechanism other than hotspot geometry contributes to the harmonic
content of the non-burst pulsations.  

The use of the daily average accretion fractional amplitude in this
analysis seemed justified because it changes smoothly (see Figure \ref{pers}).
To check whether our results were sensitive to this choice, however,
we repeated the analysis
using different time windows to compute accretion fractional
amplitudes.  We tried periods centered on the burst
time and periods of several hours immediately prior to each
burst. There was no effect on our findings; the hypothesis is still
ruled out at a level greater than one part in $10^4$.  

Figures \ref{funlum} and \ref{firlum} show that fractional
amplitudes of both the fundamental and first harmonic seem to be
relatively independent of both burst flux and rise time. Two 
features merit comment.  The first is the very low fractional
amplitude for the burst with the highest peak flux, Burst 28.  As discussed by
\citet{str03}, this burst shows evidence for photospheric radius
expansion, which would have the effect of suppressing fractional
amplitude.   The second point is more speculative. The set of fainter
bursts includes three with rather low fundamental fractional amplitudes
compared to the rest of the population. This could be due to
statistical effects, since the uncertainties on fractional amplitudes
for the fainter bursts are higher.  However, we may be seeing a sample
of a population that has a genuinely lower fractional amplitude:  it
would be interesting to look for more bursts of this type if J1814 goes
into outburst again.

\begin{figure*}
\centering
\includegraphics[width=18cm, clip]{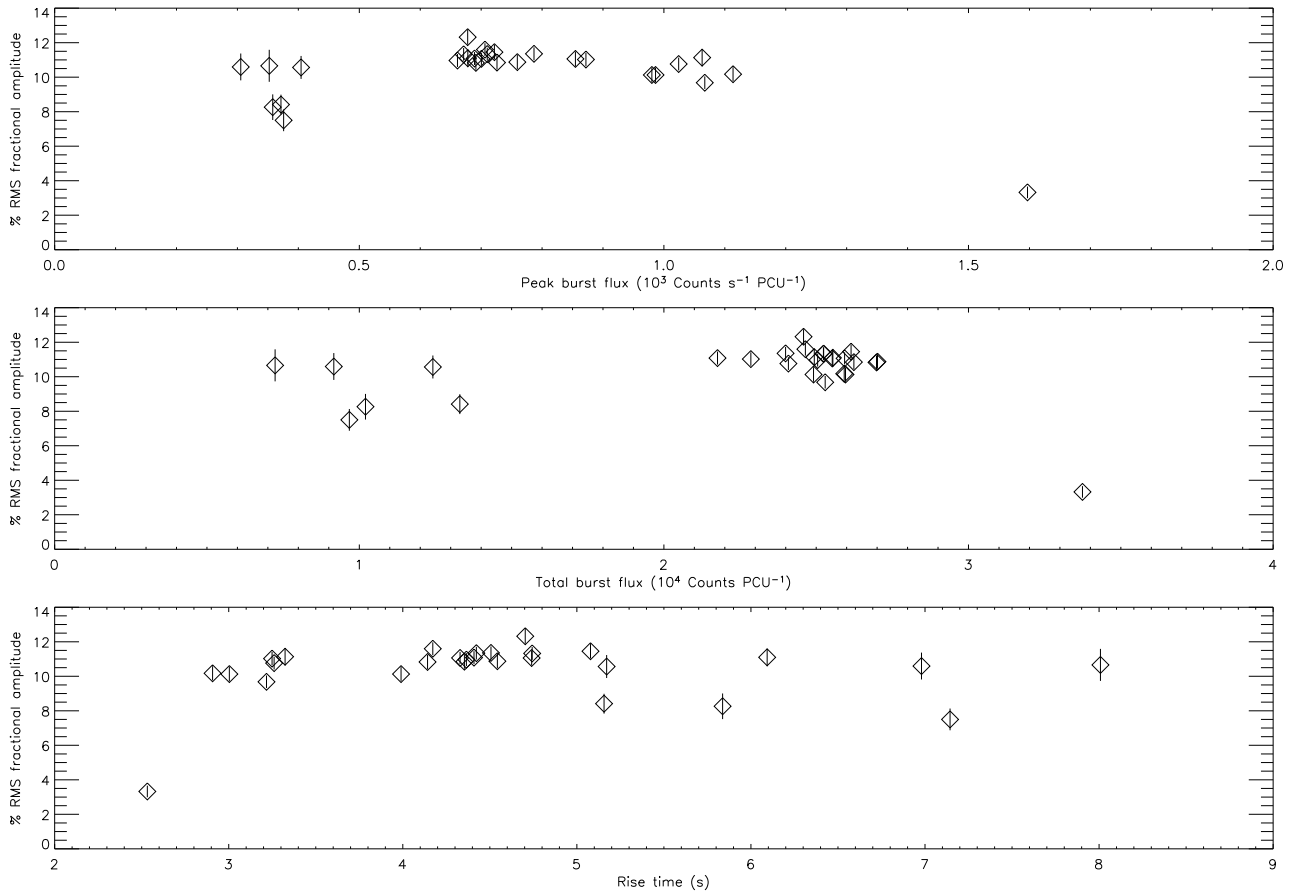}
\caption{The upper panel shows accretion-corrected fundamental
  fractional amplitude of the 
bursts against peak
burst flux (corrected for background and persistent emission).  The center
panel shows the 
behavior against total burst flux (corrected for background and
persistent emission). The lower panel shows the behavior against rise
  time, defined as the time it
  takes the corrected count rate to rise from 10\% to 90\% of the peak
  value.  }  
\label{funlum}
\end{figure*}

\begin{figure*}
\centering
\includegraphics[width=18cm, clip]{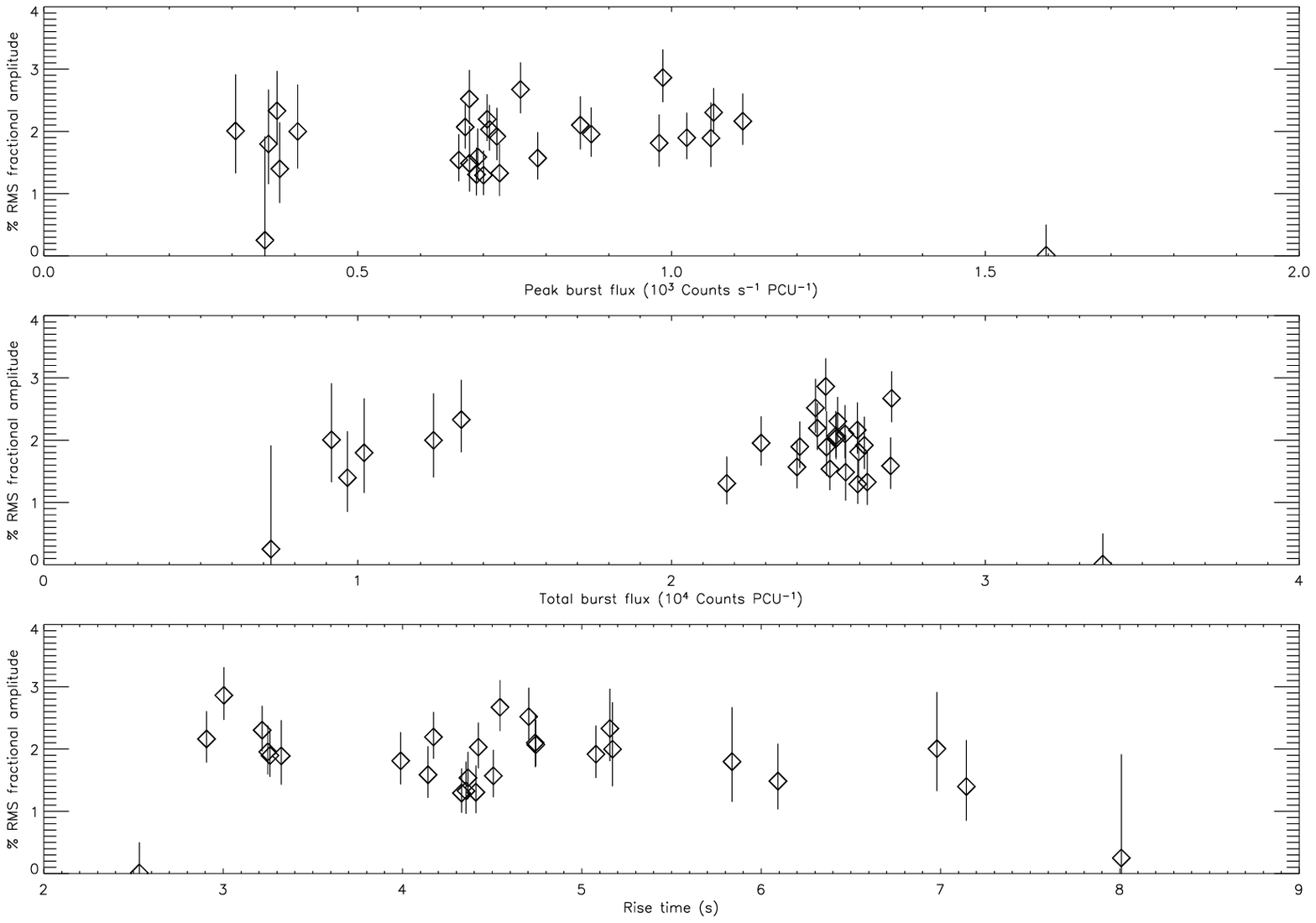}
\caption{The upper panel shows accretion-corrected first harmonic fractional amplitude
of the
bursts against peak
burst flux (corrected for background and persistent emission).  The center
panel shows the 
behavior against total burst flux (corrected for background and persistent emission). The lower panel shows the behavior against rise
  time, defined as the time it
  takes the corrected count rate to rise from 10\% to 90\% of the peak
  value.  } 
\label{firlum}
\end{figure*}

\section{Variability during bursts}
\label{ibv}

\subsection{Variation in fractional amplitude}
\label{ibv1}

\citet{str03} noted that the bursts of J1814 seemed to show
variations in rms fractional amplitude of $\approx 3-5$ \%, on
timescales of 7 to 15 s.  In this section we present an in-depth
analysis of this variability.  We focus only on the behavior of the
fundamental; the first harmonic is sufficiently weak
that we cannot hope to make reasonable detections on shorter
timescales than those analysed in section \ref{obv}.  

The upper panels of Figure \ref{b1} show the light curves
and fractional amplitude
variations for a sample of the bursts.  The burst average fractional
amplitude is also shown for comparison.
Fractional amplitudes are calculated only when the flux is at least
twice the pre-burst level (to reduce the contribution of the accretion
component to any variability).  We use non-overlapping bins of 5000
photons.  The durations of these bins are sufficiently short that we
assume frequency to be constant during each bin, and select the
frequency $\nu_m$ for which $Z_m$ is maximised.  The effect of this assumption
on the distribution of $Z_m$ values is discussed below.  

\begin{figure*}
\centering
\includegraphics[width=8.5cm,clip]{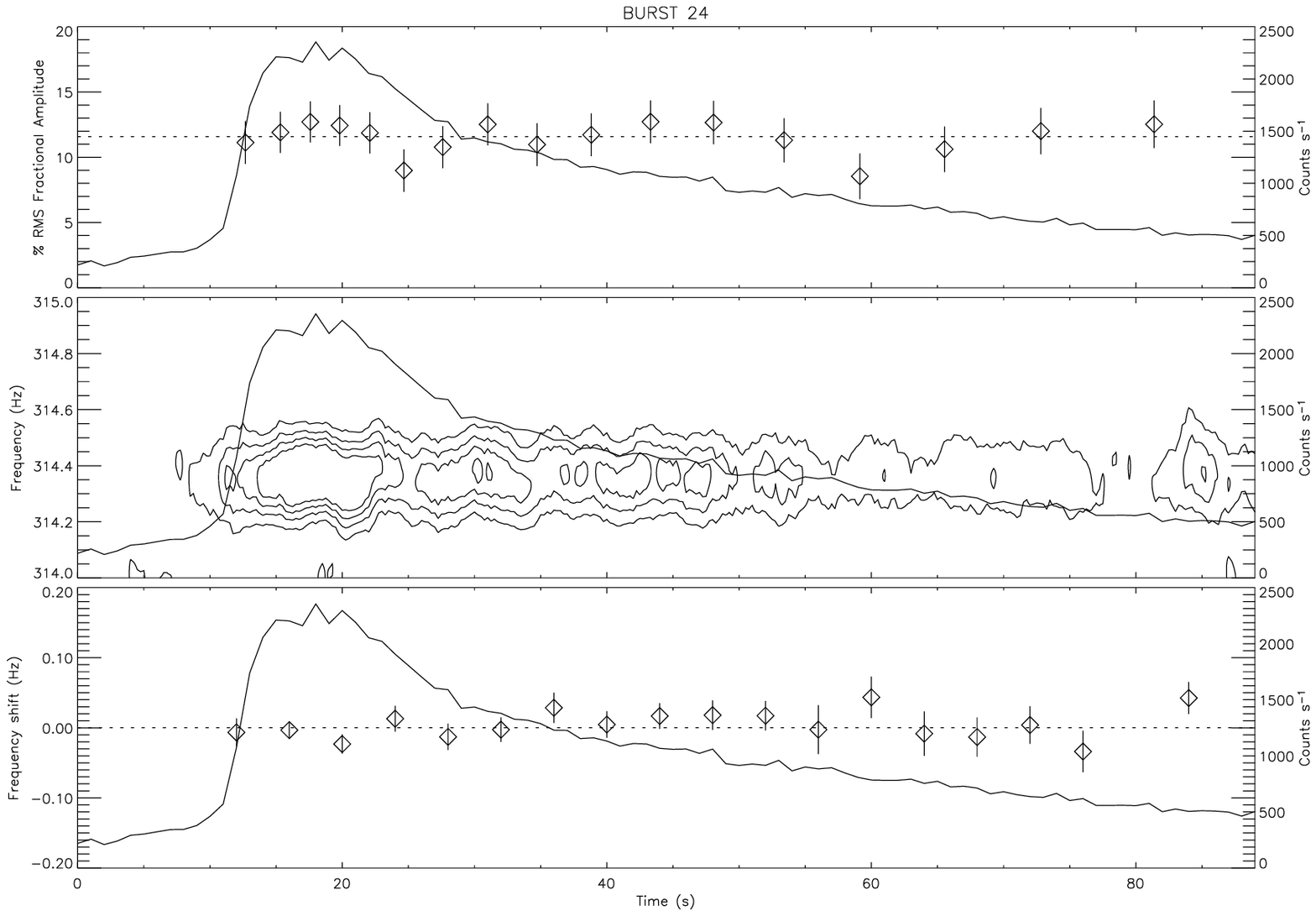}
\includegraphics[width=8.5cm,clip]{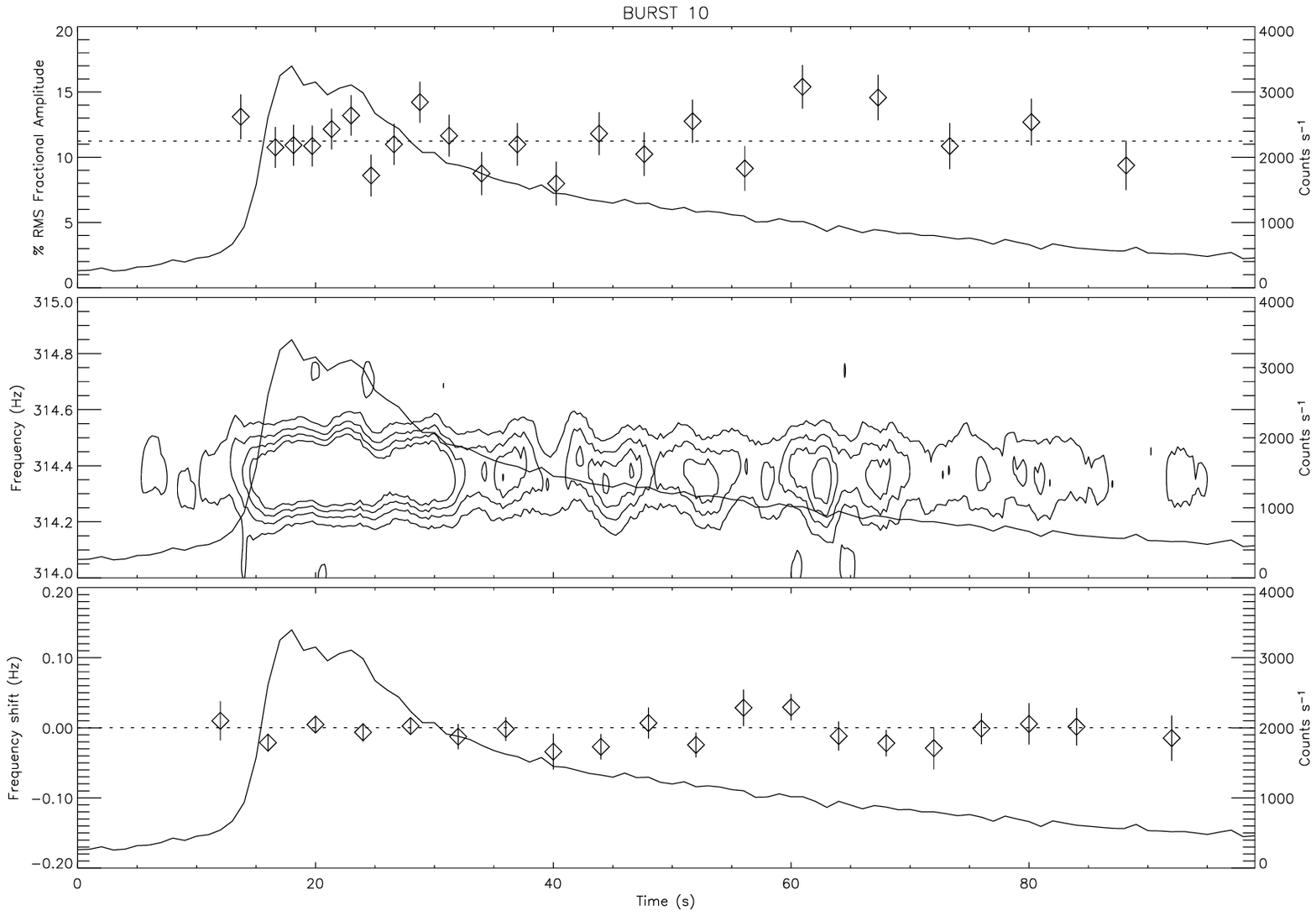}
\includegraphics[width=8.5cm,clip]{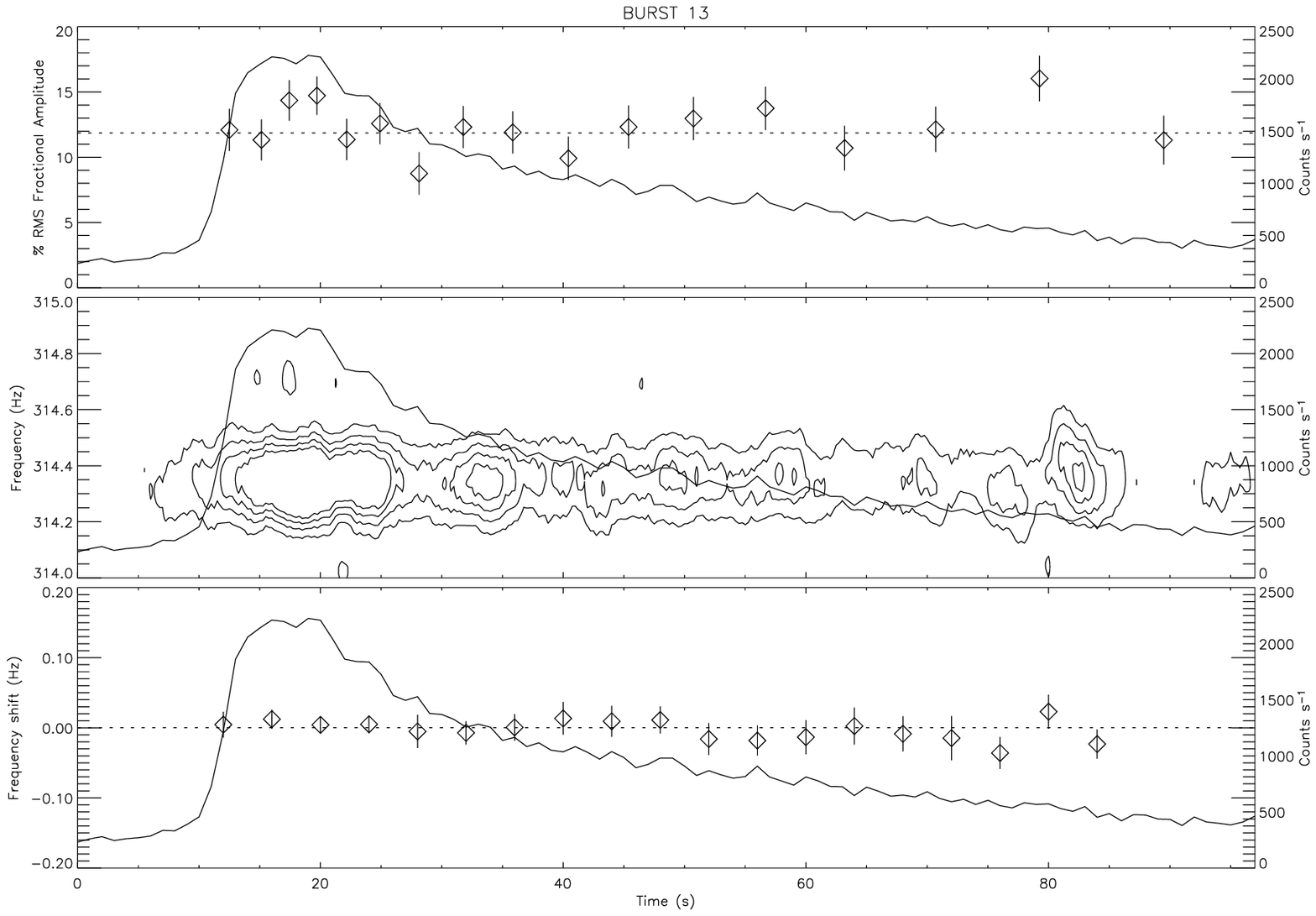}
\includegraphics[width=8.5cm,clip]{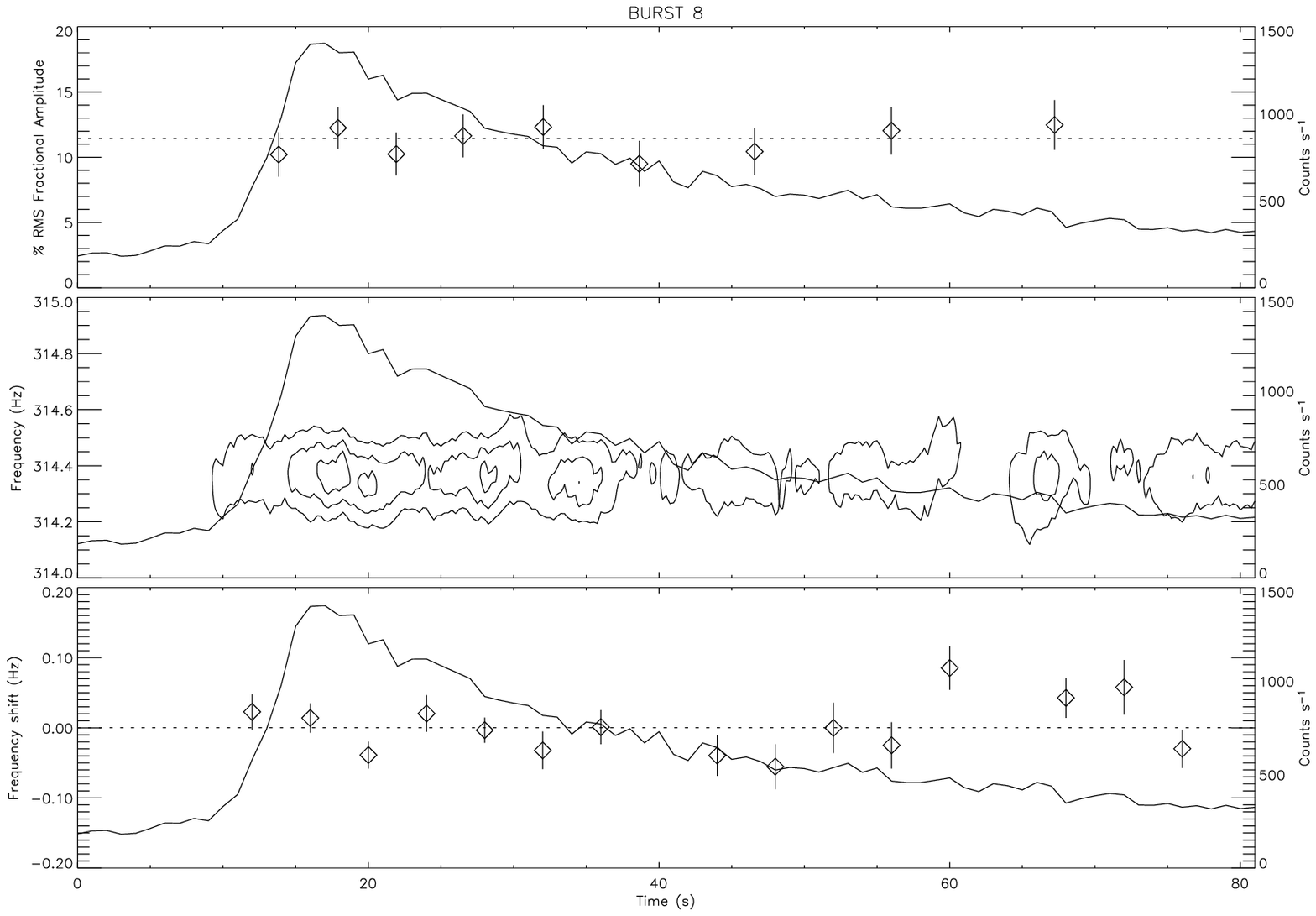}
\includegraphics[width=8.5cm,clip]{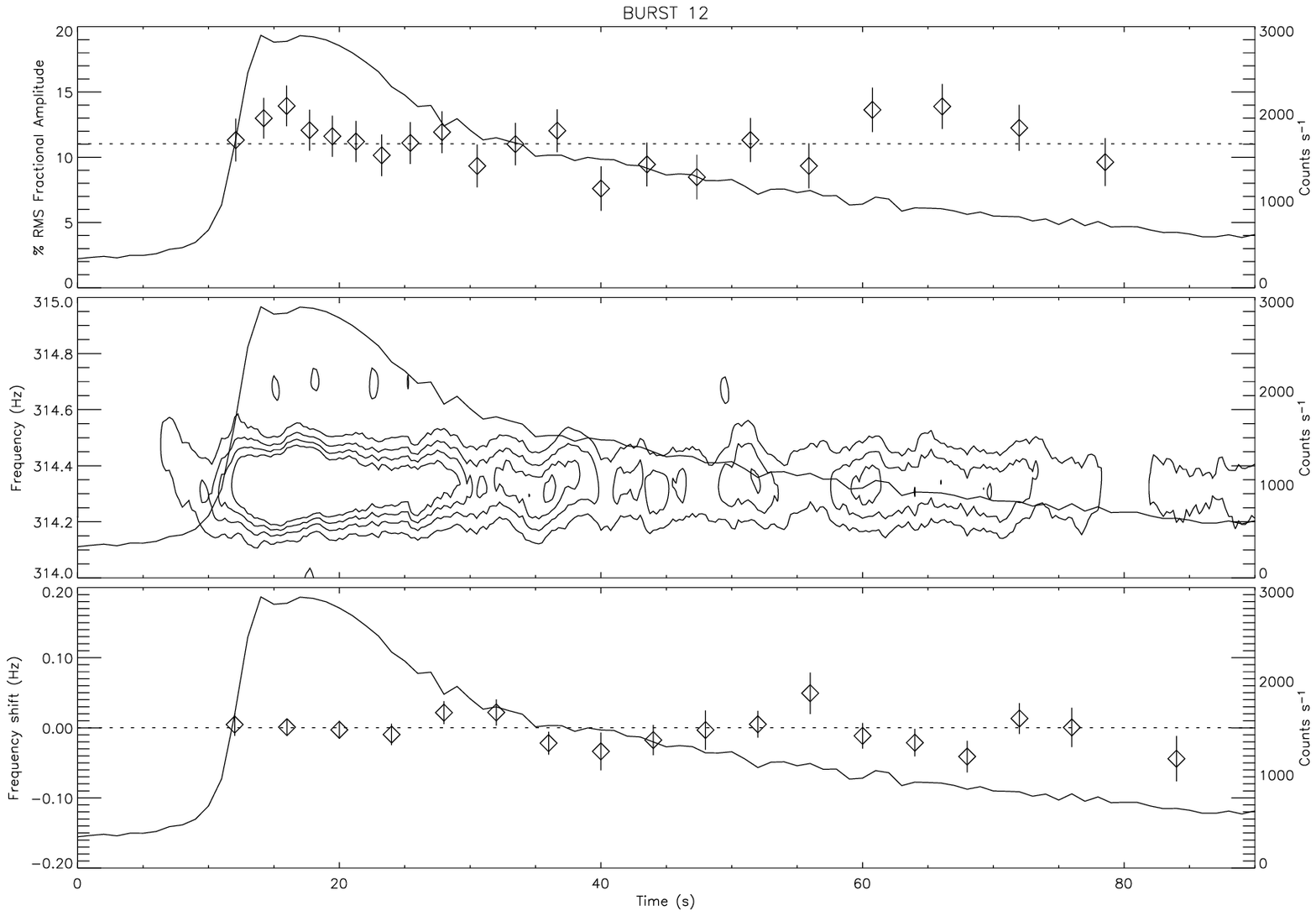}
\includegraphics[width=8.5cm,clip]{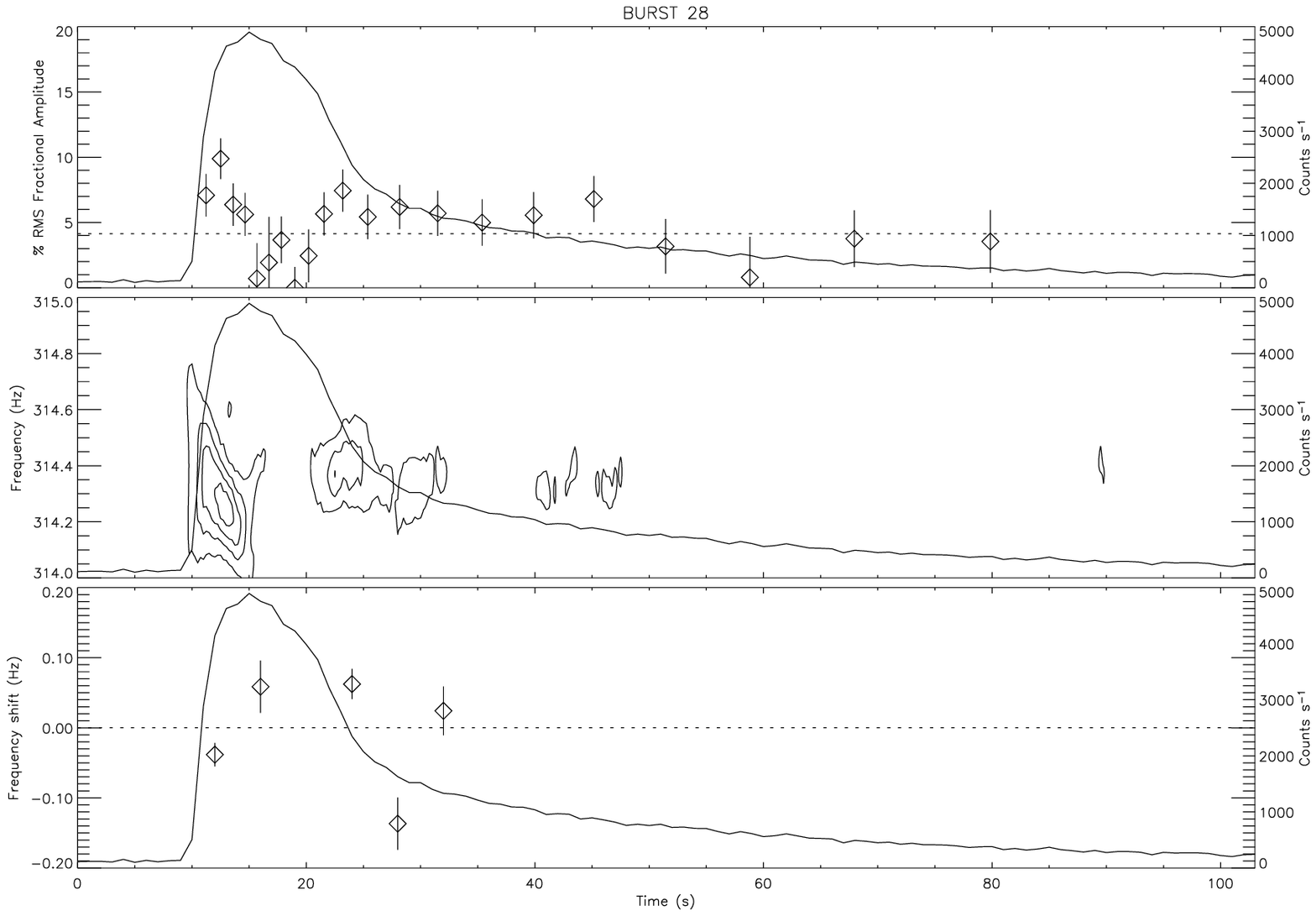}

\caption{Light curves for a sample of the bursts (shown as solid lines).  Upper
  panel: fractional amplitude   
  of the fundamental, calculated for non-overlapping bins of 5000
  photons. The dotted line indicates burst average fractional
  amplitude. Center panel:   Dynamical power spectra, calculated using
  overlapping 4s bins, with new bins starting at 0.25s intervals.  The
  contours show $Z^2_1 = 15, 35, 55, 75$.  Lower panel:  The
  difference between peak frequency (calculated for
  non-overlapping 4s bins and a minimum signal strength $Z_m=15$) and
  best-fit orbital frequency.  The error bars shown in the plots
  are the equivalent of 1$\sigma$ error bars:  there is further
  discussion of error bars elsewhere in the paper.   Burst 24 has the
  lowest amplitude variability; Burst 10 one of the highest.  Burst 13 has
  the lowest frequency variability; Burst 8 one of the
  highest.  Bursts 12 and 28 are discussed in more detail in the main
  text; note that Burst 28 has the highest
  frequency and fractional amplitude variability of any of the bursts.}
\label{b1}
\end{figure*}

The fractional amplitude during the bursts does seem to vary
around the burst average value, with the difference between maximum
and minimum fractional amplitudes measured during a burst lying in the range
2.4 to 9.9\%. As in Section \ref{obv1}, we want to know whether the observed
variability is consistent with that which we might expect due to
statistical effects 
alone.  The null hypothesis is that fractional
amplitude remains constant over the course of a burst.  To quantify
the variability we evaluate the following quantity for 
each burst:

\begin{equation}
\bar{\chi}^2 = \sum_i^{N_\mathrm{bins}} (r_i - r_\mathrm{model})^2
\end{equation}
$N_\mathrm{bins}$ is the number of non-overlapping 5000 photon bins for a
given burst and $r_i$ is 
the fractional amplitude calculated for that bin.  In this case
$r_\mathrm{model}$ is a constant that is varied until $\bar{\chi}^2$ is minimised.  

To simulate the distribution of $\bar{\chi}^2$ if the hypothesis is
true, in order to estimate 
the significance of the measured value, we proceed as follows. For
each burst we fit the light curve,  
and generate simulated events files, imposing a constant fractional
amplitude oscillation on the light curve.  The fractional amplitude
assumed for each burst is the burst average fractional amplitude
calculated in the previous section and listed in Table \ref{bdatatab}.  We then
analyse the simulated events files in exactly the same way as we
analyse the real data, taking 5000 photon segments and calculating a
$Z_m$ for each segment.  We then compute the value of $Z_s$ for which
$f(Z_m:Z_s)=0.5$ and use this to calculate a fractional amplitude for
the segment.  We run 1000 simulations for each burst and compute the
quantity $\bar{\chi}^2$ for each simulations.  For each burst we can
then determine the probability $V_r$ that the simulated $\bar{\chi}^2$
exceeds the measured value.  This gives us a simple measure
of the variability of each burst.  The results are given in Table
\ref{bdatatab} and Figure \ref{burstvarb}.  Note that for Burst 28 we
ran additional ($10^4$) simulations to confirm the high variability of
this burst.  

\begin{figure}
\centering
\includegraphics[width=8.5cm,clip]{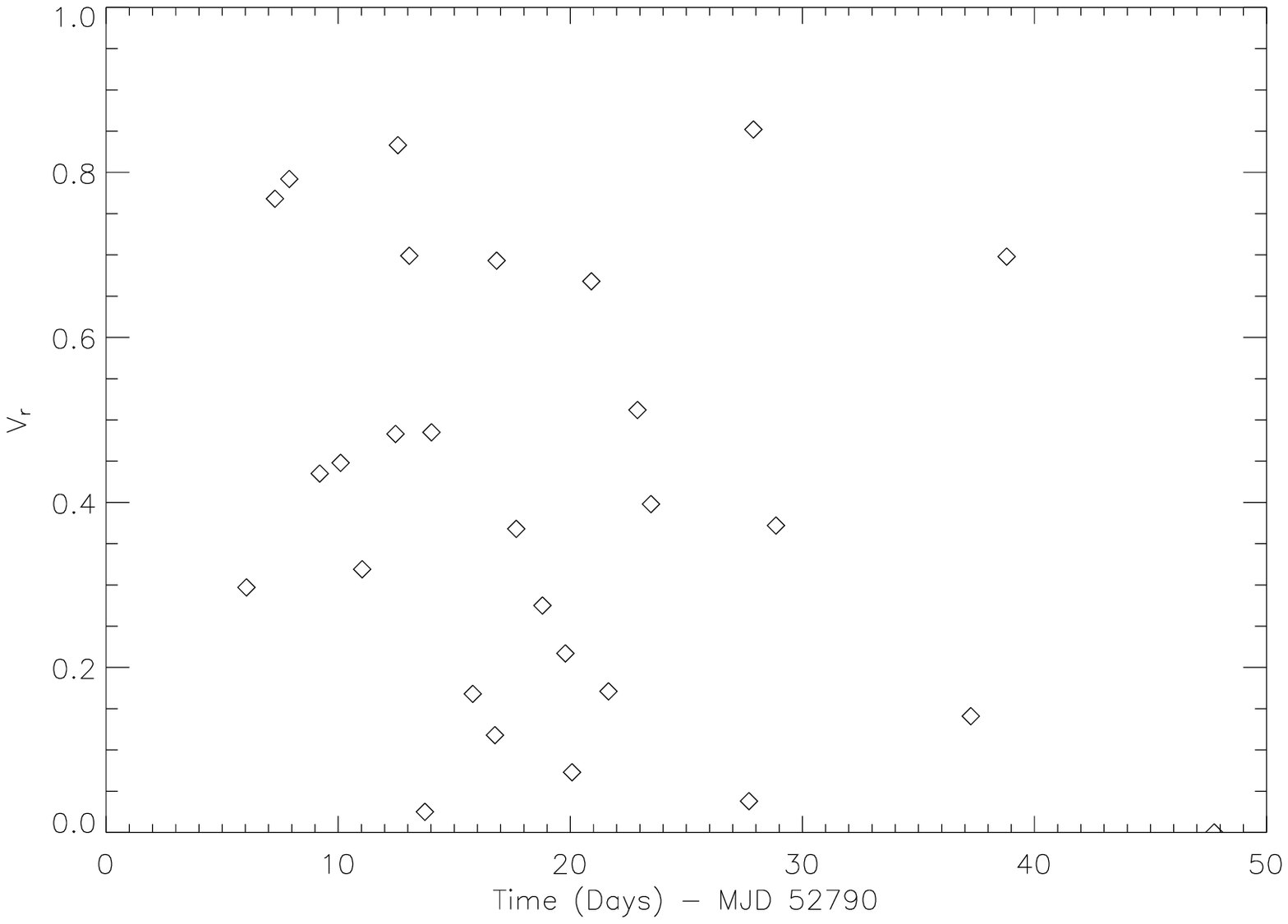}
\caption{Fractional amplitude variability of each burst as quantified
  by $V_r$.  Low
  values indicate highly variable bursts, high values low 
  variability bursts.  Note that there are no bursts for which
  $V_r$
  exceeds 0.852.  If the null (constant fractional amplitude)
  hypothesis were true one would expect about 4 bursts in this range.
  Nonetheless the deviation is not sufficiently significant to rule
  out the hypothesis.
}
\label{burstvarb}
\end{figure}

A legitimate question is why we used simulated light curves to
generate the distribution of fractional amplitudes, rather than
adopting the approach that we took in Section \ref{obv}.  We do this
because we need to take into
consideration the effect 
of allowing a degree of freedom in frequency selection.  Choosing the
frequency that maximises $Z_m$ has the effect of skewing $p(Z_m:Z_s)$
away from the theoretical distribution.  The effect is not large, but
the peak of the distribution moves to a value of $Z_m$ that is
a few units higher than the theoretical prediction.  That this is to
be expected can be seen by considering the distribution in the absence
of any signal.  Running Monte Carlo simulations and measuring $Z_m$ at
a fixed frequency we find that the noise  powers
are distributed around the expected value $Z_m=2$ (for one harmonic).  If
however we consider a range of frequencies and always pick the
maximum $Z_m$ within that range,  it is clear that we will shift the
peak of the 
distribution of noise powers to a higher value.  Because of this
skewing effect, we choose to use Monte Carlo simulations of the light
curve to generate the distribution of fractional amplitudes rather than relying on
the theoretical distribution.  This issue did not affect the
analysis in Section \ref{obv} because we used a
particular frequency model and did not skew the model to maximise
$Z_m$.

Having established how variable each burst is, we now need take into
account the fact that we have  a number of bursts. Each burst is an
equally valid test of the hypothesis that fractional 
amplitude remains constant during a burst.  We therefore need to know
the expected distribution of burst variabilities.  Fortunately because
we are using significances to measure variability the theoretical
distribution of variabilities is very simple:  the set of values of
$V_r$ should be 
uniformly distributed between 0 and 1.  This means that we expect to
see a range of burst variability.  

The first thing that is clear is that even though we have nearly 30
bursts, Burst 28 constitutes a highly significant outlier.  The
fractional amplitude for this burst is without doubt varying.  The
most probable explanation for this is that photospheric radius
expansion in the peak of the burst obscures the stellar surface,
suppressing the fractional amplitude until touchdown.  

What about the first 27 bursts?  We will consider these bursts as a
set and assume that the physical processes occurring in each burst are
the same (Burst 28 is excluded from the set because there is an
additional physical process, photospheric radius expansion, at work in
this burst).  Figure \ref{ampvardist} shows that there is a
shift in the distribution towards lower values of $V_r$, suggesting
that there may be some variability.  However,
we need to evaluate formally the significance of the deviation.  There
are various  ways of doing this but given that the deviation takes the
form of a shift, a simple option is to
compute the mean of the 27 values of $V_r$.  We can then use simulations to model the
distribution of means (as one would expect these values are
distributed symmetrically about 0.5) and use this distribution to
measure the significance of the measured value.  We find that the mean
of the dataset is 0.420.  However, values smaller than this are obtained in 74
out of 1000 simulations.  The hypothesis that fractional amplitude
remains constant during the bursts cannot be rejected.

\begin{figure}
\centering
\includegraphics[width=8.5cm,clip]{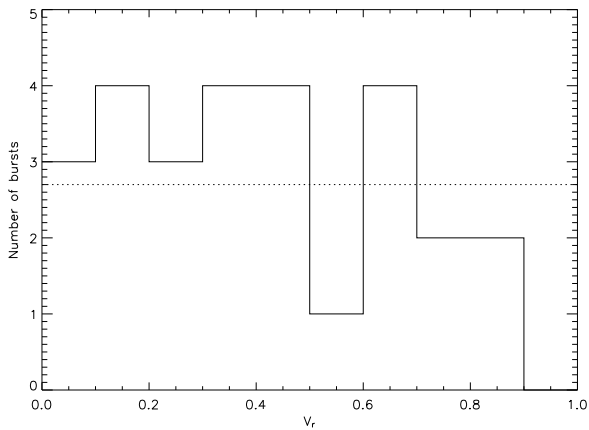}
\caption{Histogram showing the distribution of the $V_r$ values for
  the first 27 bursts.  The dashed line indicates the distribution
  that we would expect if there were no variation in fractional
  amplitude during the bursts. The histogram shows that there seems to
  be a shift towards lower values of $V_r$, suggesting some
  variability in fractional amplitude; however the deviation is not 
  sufficiently significant to rule 
  out the constant fractional amplitude hypothesis.}
\label{ampvardist}
\end{figure}

Similar results are obtained if one uses a Kolmogorov-Smirnov (KS) test,
a standard technique for measuring deviations from a theoretical
distribution.  The KS test is supposed to be sensitive to shifts in
distribution but should not be used if there are major outliers in the
data set (such as Burst 28).  The KS test indicates that the
probability of obtaining 
a deviation larger than the observed value and of the same sign as the
observed value is 0.12.  Again, the hypothesis cannot be rejected.   

In the above analysis we have computed a measure of the variability of
each burst and then analysed the distribution of burst
variabilities.  This method treats each burst as an equally valid test
of the hypothesis that fractional amplitude remains constant during a
burst.    An alternative way of analysing the
data is not to treat each burst 
separately, but to compute a rolling $\bar{\chi}^2$ where we sum over
all of the 5000 photon data points from all of the bursts.  If we do
this we find that the simulated $\bar{\chi}^2$ exceeds the measured
value in 69 simulations out of 1000, in good agreement with the
previous results.  Note
however that by rolling all of the results together we can no longer
distinguish between a set of bursts that all have medium variability,
and a set where the variability ranges from low to high.  The
first option would in fact be rather unusual, as it would imply the
outliers were evenly distributed throughout the data set, whereas
groupings of outliers will occur by chance. We can distinguish this
behavior only by treating each burst individually as we did in our
initial analysis.  Nonetheless, the 
combined test does (as we would expect) back up our conclusion that we
cannot rule out the constant fractional amplitude hypothesis.

Note that in this
analysis we have not separated out possible 
variability in the accretion component from possible variability in
the thermonuclear component.   In addition we did not take into
account the temporal ordering of data points during a burst.  This could obscure
interesting effects such as, for example, the peak fractional
amplitude always occurring at the peak of the lightcurve.  To test for
any such effects we proceed as follows.  We first convert the
fractional amplitude values evaluated during each burst into a
percentage shift above or below the burst average value.  We then take
the full set of data for the first 27 bursts, bin the data
points by time elapsed since the start of the burst, and compute the
mean percentage shift in fractional amplitude for each time bin.  For
a range of different time bin sizes, we find no statistically significant
deviation from zero:  we cannot detect any consistent temporal effect.
In particular, we  
find no evidence for very high fractional amplitudes very early in
the burst rise.    This effect, seen in the brighter bursts of some
LMXBs, is thought to be caused 
by the rapid spreading of the initial ignition hotspot \citep{str97,
  str98}.  Assuming 
ignition does occur at a point, a similar effect should occur in
J1814.  Detecting such an effect would however be difficult for two
reasons:  firstly, the bursts of J1814 are faint; and secondly because
the
effect would be masked by the accretion-related pulsations.  

\subsection{Variation in frequency}
\label{ibv2}
 
Let us now move on to consider variability in frequency.   The center
panels of  Figure \ref{b1} show dynamical power spectra computed using
overlapping 4 s bins, with new bins being started at intervals of
0.25 s.  The lower panels of Figure \ref{b1} show the
difference between the  frequency $\nu_m$ at which $Z_m$ is maximised
and the frequency derived from the best 
fit orbital model, for non-overlapping 4s bins.  The error
bars on the frequency measurements are derived from simulations, as
discussed below.  

Monte Carlo simulations show that the difference between the measured
peak frequency and the input model frequency depend on both the length
of the time bin considered and on the value of $Z_m$.  For this reason
we fix the time bin size.  The distribution of apparent frequency shifts
with $Z_m$, for a fixed binsize of 4s, is shown in Figure \ref{fdist}.
It is clear that below a certain value of $Z_m$
it is no longer possible to define even a 1$\sigma$ error bar.  We
therefore set a minimum $Z_m$, below which we will not attempt to
compute frequency.  We set a minimum of $Z_m = 15$, the lowest level
at which we can still compute a 2$\sigma$ error bar.  Note that a
similar problem afflicts phase residuals; below a certain signal
strength the uncertainty in the profile fit is sufficiently large that
one cannot define error bars on the phase.

\begin{figure}
\centering
\includegraphics[width=8.5cm, clip]{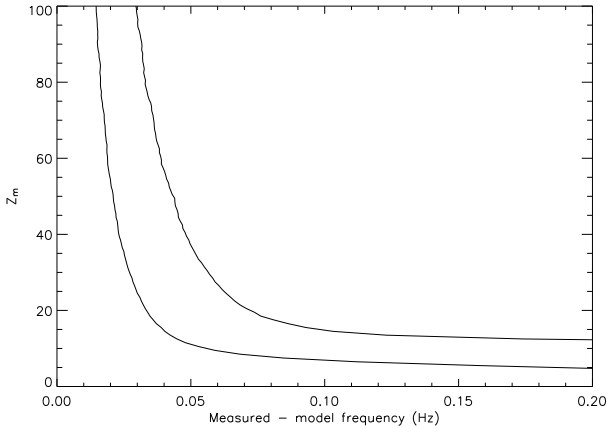}
\caption{Frequency distributions derived from Monte Carlo
  simulations, for independent time bins of 4s duration. The lines
  shown are 1$\sigma$ and 
  2$\sigma$ error bars.} 
\label{fdist}
\end{figure}

The results in the lower panels of Figure \ref{b1} are
computed using only events that arrive when the flux is at least twice the
pre-burst level. We will discuss frequency shifts in the burst rise
later in this section.  The figures suggest that there is at most only
low variability in frequency during the bulk of
the burst.  Let us now quantify this more precisely.  The null
hypothesis that we wish to test is that the underlying frequency 
remains constant, apart from any orbital corrections, during the main
part of the burst (excluding the earliest stages of the burst rise).
We define the following simple measure of frequency variability:

\begin{equation}
\bar{\chi}^2_\nu = \sum_i^{N_\mathrm{bins}} (\nu_i - \nu_\mathrm{model})^2
\end{equation}
where $N_\mathrm{bins}$ is the number of non-overlapping 4s bins in
the burst (where flux is at least twice the pre-burst level), $\nu_i$
is the $\nu_m$ calculated for each bin, and $\nu_\mathrm{model}$ is in
this case the frequency derived from the orbital ephemeris.  The
decision to use non-overlapping bins is motivated by the fact that
the dynamical power spectra do not show evidence for significant
variations in frequency on timescales shorter than 4s apart from
perhaps in the burst rise phases, which we treat in more detail later
in this section.  

As before we use Monte Carlo simulations to determine the
significance of the $\bar{\chi}^2_\nu$ measured for each
burst. However, the approach that we used previously, simulating the
light curve to generate artificial events files, is no longer
appropriate.  This is because the distribution of frequency shifts
depends on $Z_m$ and hence on fractional amplitude.  In the previous
section we have shown that we cannot rule out the hypothesis that 
fractional amplitude remains constant during a burst.  However, we
have not tested any models that call for variation (to the best of our
knowledge there are no quantitative predictions in the literature at
this time).  Thus we cannot exclude the
possibility that a model with underlying variation might fit the
data better.  For this reason we choose to be cautious and make no
assumptions about the underlying behavior of the fractional
amplitude.  However, without a model for the fractional amplitude
evolution we cannot generate simulated
lightcurves in order to compute the distribution of
$\bar{\chi}^2_\nu$.

Instead we use the simulated distributions of frequency shift
as a function of $Z_m$, $f(\nu_m-\nu_\mathrm{model}:Z_m)$,  that were used to
generate Figure \ref{fdist}.  For each burst we have a set of
$N_\mathrm{bins}$ values of $Z_m$. If we are running $N_\mathrm{sim}$
simulations we therefore start by generating $N_\mathrm{sim} \times
N_\mathrm{bins}$ random numbers drawn from a uniform distribution
between 0 and 1. For each data point in the burst we take
$N_\mathrm{sim}$ of these $f$ values and read off the corresponding
frequency shifts from the simulated cumulative distribution function
for the appropriate value of $Z_m$.  This process is repeated for each
data point in the burst.  We then group the simulated frequency shifts
into sets of $N_\mathrm{bins}$, each set containing one point from
each $Z_m$ value, and compute a $\bar{\chi}^2_\nu$ for that set.  For
each burst we can 
then determine the probability $V_\nu$ that the simulated $\bar{\chi}^2_\nu$
exceeds the measured value.  This gives us a simple measure
of the variability of each burst.  The results are given in Table
\ref{bdatatab} and Figure \ref{fvar}.  It is readily apparent that
none of the bursts are major 
outliers in the way that Burst 28 was when we computed $V_r$.

\begin{figure}
\centering
\includegraphics[width=8.5cm, clip]{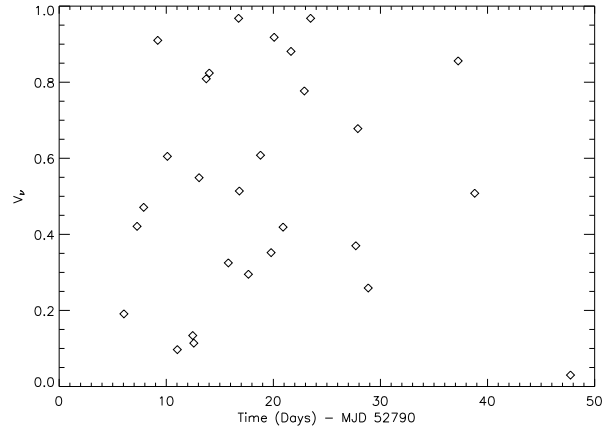}
\caption{Frequency variability of each burst as measured
  by $V_\nu$. Low values indicate highly variable bursts, high values low
  variability bursts. }
\label{fvar}
\end{figure}

As before we now need to take into account the fact that we have 28
bursts.  This time we include Burst 28 in our sample because
photospheric radius expansion only affects frequency variability in
the sense that it suppresses $Z_m$, and we have taken this into
account in our simulations.  The expected distribution of $V_\nu$,
assuming that there are no shifts beyond orbital corrections, should
be uniform.  Figure \ref{freqvardist} illustrates that there is
 good agreement between the theoretical distribution and the
observed distribution, and certainly no skew towards lower values of
$V_\nu$, as we would expect if there is significant frequency
variability. This is confirmed by the fact that the mean of the set of
values of $V_\nu$ is 0.530.  A KS test also confirms no significant
deviation of the distribution uniformity.  In addition if we combine 
 data points from all of the bursts and compute a rolling
 $\bar{\chi}^2_\nu$, we find that the measured value is exceeded in 616
 out of 1000 simulations.  There is therefore no evidence of frequency
 variability during the main part of the bursts.  

\begin{figure}
\centering
\includegraphics[width=8.5cm,clip]{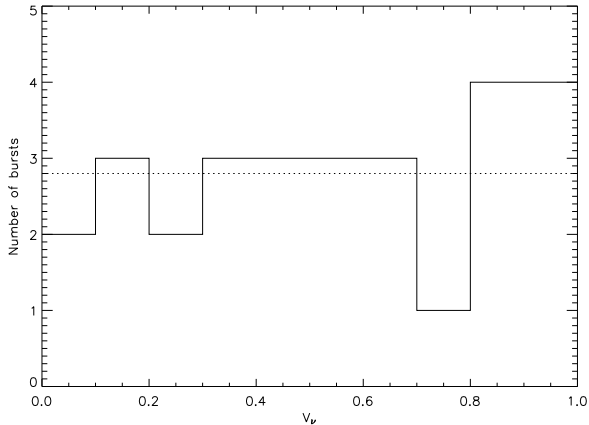}
\caption{Histogram showing the distribution of the $V_\nu$ values for
  all 28 bursts.  The dashed line indicates the distribution
  that we would expect if there were no variation in frequency during
  the bursts. There is no skew towards lower values of $V_\nu$, as we
  would expect if there was frequency variability.}
\label{freqvardist}
\end{figure}

We also searched for any statistically significant temporal ordering
effects. Such an effect might be the minimum frequency always
occurring at the peak of the lightcurve, which could indicate a
statistically significant frequency decrease at this time.  By
combining measurements from all of the different bursts and binning by
time elapsed since burst start we have verified that we detect no
significant temporal ordering effects in the frequency variability.  

The analysis
presented thus far has focused on events arriving when the
flux is at least twice the pre-burst flux. Let us now consider the
burst rise in more detail, by including events that occur before this
flux threshold is reached.  \citet{str03} noted that
for two of the bursts (Bursts 12 and 28; note the difference in
burst numbering compared to \citet{str03}), there appeared to be 
significant frequency decreases during the burst rise. These two
apparent shifts, visible in Figure \ref{b1}, are
sufficiently rapid that we must use either overlapping time bins or shorter time bins to resolve  
them.  

The difficulty with measuring frequency shifts in the burst rise is
that in general, the signal ($Z_m$) is weak, which can lead to large
statistical variations. If we use shorter, say 2s time bins, this
increases the statistical scatter of the frequency measurement,
particularly for weak signals (effectively broadening the error bars
shown in Figure \ref{fdist}).  Overlapping time bins also have their
problems, however, as they give rise to correlated noise.  

The dynamical power spectrum and phase residual evolution for the rise
of Burst 12 are shown in Figure \ref{b12}.  The power spectrum
suggests a frequency very close to the orbital frequency before
the burst.  The frequency then jumps a few tenths of a Hz
above the orbital frequency and gradually slides back down to the
orbital frequency before the burst rise is complete.  We must
determine whether the jump and subsequent slower drop in the frequency
are statistically significant. We will argue that they are not.

\begin{figure*}
\centering
\includegraphics[width=18cm,clip]{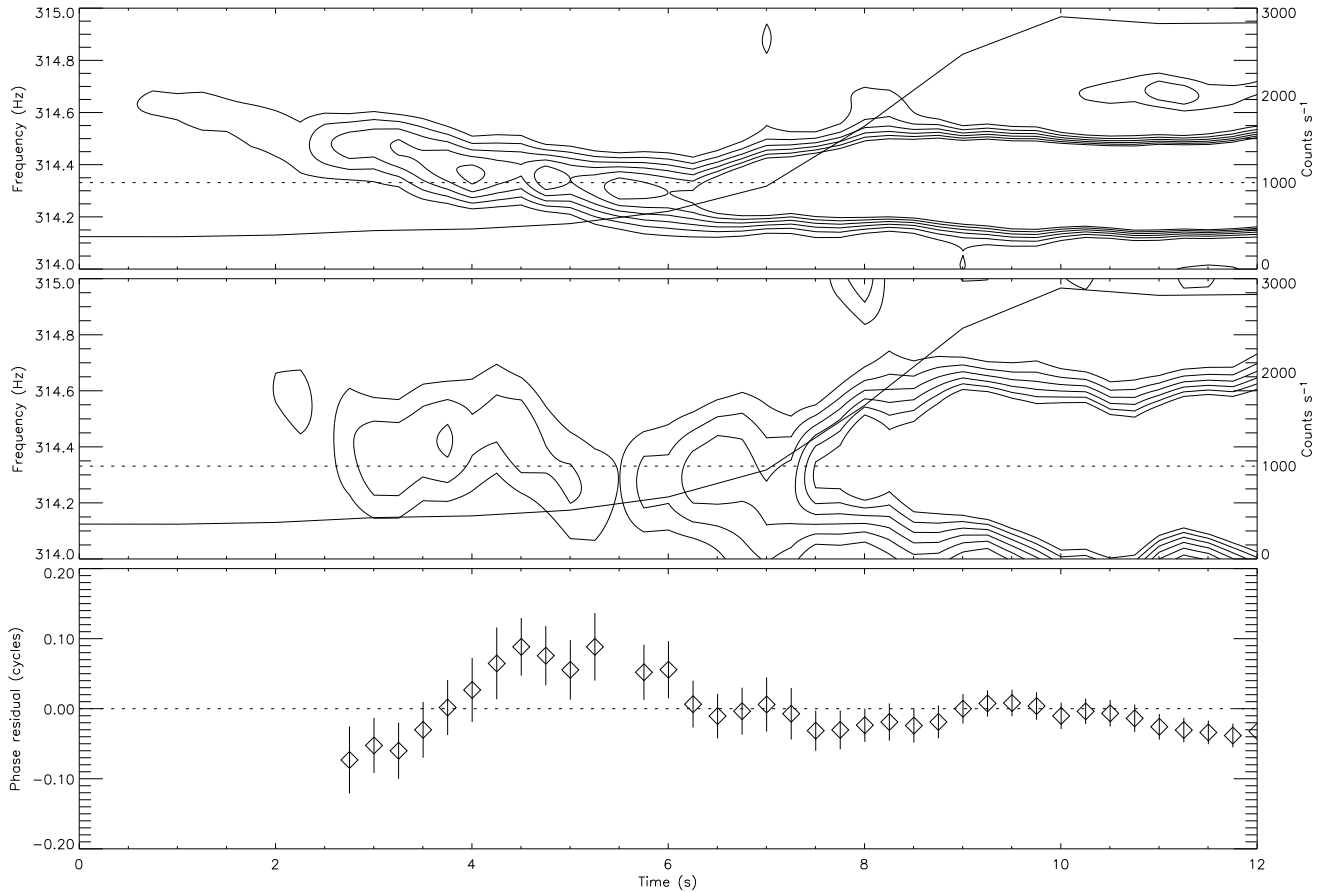}
\caption{The burst rise of Burst 12.  The upper panel shows the
  dynamical power spectrum for 4s bins, with a new bin starting every
  0.25s.  The center panel shows the dynamical power spectrum for 2s
  bins, with a new bin starting every 0.25s.  The contour levels are
  $Z_m = 10, 15, 20, 25, 30, 35$.  The dotted lines in the upper and
  center panels indicate the best-fit frequency according to the
  orbital model. The lower panel shows phase
  residuals compared to the best fit orbital frequency evolution
  model, for 2s bins with new bins starting every 0.25s.   }
\label{b12}
\end{figure*}

The first point after the jump has $Z_m \approx 10$.  For such a weak
signal obtaining an apparent frequency shift of the observed magnitude
or larger is possible in $\approx 10$\% of simulations.  Given that we have 28
bursts, therefore, this is not unusual. The events
that gave rise to the initial large shift are still present in the
next 4 seconds' worth of power spectra.  Could they be acting to keep
the frequency high, given that by the next independent time bin the
frequency is back to the orbital frequency?  The answer is yes
provided that the frequency shift is
sufficiently large and the light curve climbs only slowly over the
period.  Both of these conditions are met in this burst.  

To verify whether this is the case we have re-computed the dynamical
power spectra using shorter (2s) bins, to reduce correlation effects.
Although there is still some
movement in frequency the large downshift is no longer apparent.  The
magnitude of the shift is not statistically significant given our sample size. 
Dynamical power spectra and the associated phase residuals for the 2s
bins are shown in Figure \ref{b12}.  

The dynamical power spectrum and phase residual evolution for the rise
of Burst 28 are shown in Figure \ref{b28}.  The peak frequency shifts
from the orbital frequency down by $\approx 0.15$ Hz during the burst
rise. The peak signal during this phase is strong: in the range $Z_m =
60-80$. For such a strong signal correlated noise is unlikely to play
a role, and we have confirmed this by re-computing the dynamical power
spectrum using shorter time bins.  The shift is also readily apparent
from the phase residuals. Simulations indicate that the
probability of obtaining a frequency shift this large for a signal
this strong is less than 1 in $10^4$.  Even given the number of bursts
sampled this is a significant outlier.  We can with a high degree of
confidence assert that the frequency drops below the orbital frequency in the
rise of Burst 28.

\begin{figure*}
\centering
\includegraphics[width=18cm,clip]{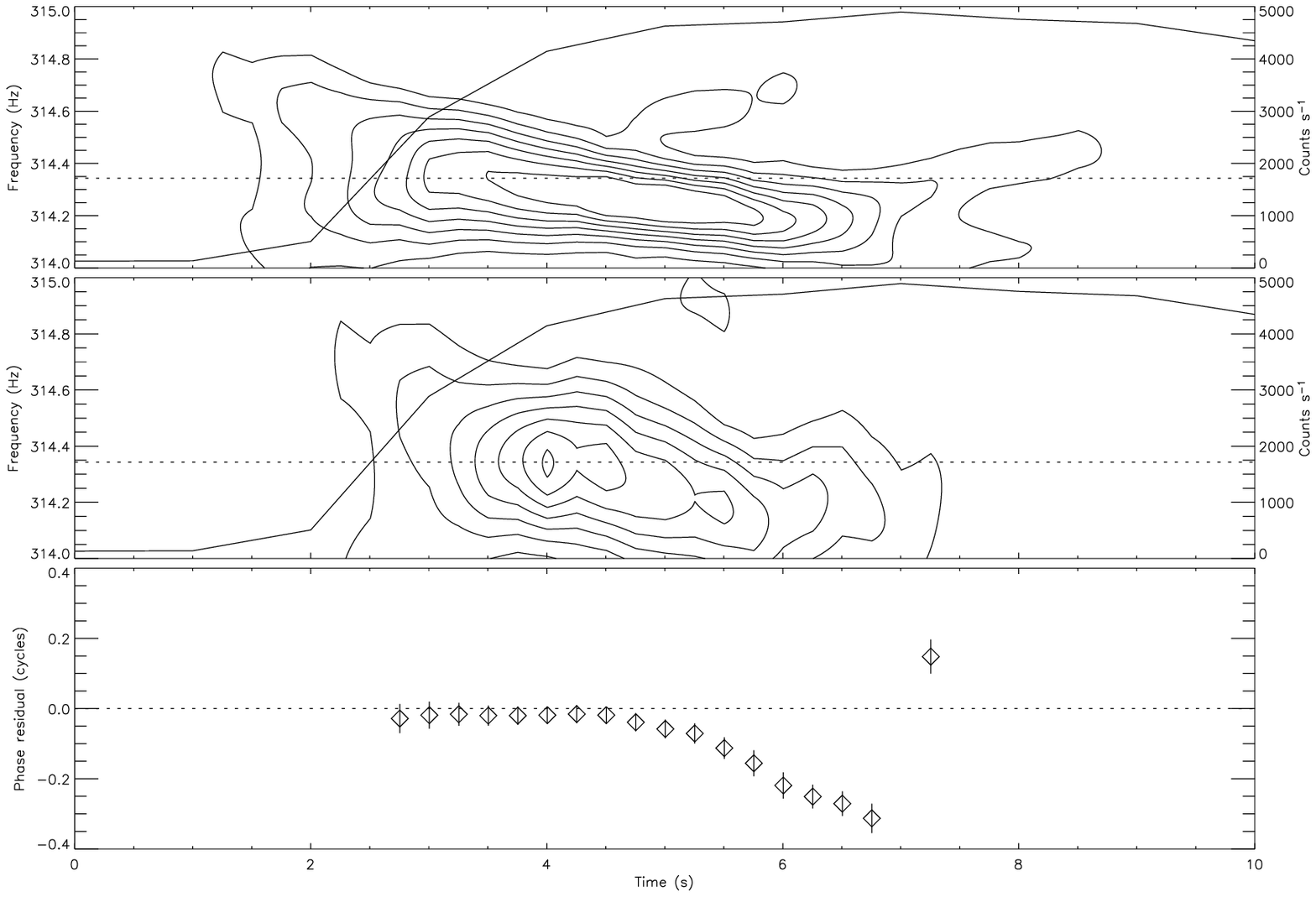}
\caption{The burst rise of Burst 28.  The upper panel shows the
  dynamical power spectrum for 4s bins, with a new bin starting every
  0.25s.  The center panel shows the dynamical power spectrum for 2s
  bins, with a new bin starting every 0.25s.  The contour levels are
  $Z_m = 10, 20,30, 40, 50, 60, 70$. The dotted lines in the upper and
  center panels indicate the best-fit frequency according to the
  orbital model. The lower panel shows phase
  residuals compared to the best fit orbital frequency evolution
  model, for 2s bins with new bins starting every 0.25s.}
\label{b28}
\end{figure*}

In summary, none of the bursts apart from Burst 28 show any evidence
for frequency shifts in the burst rise.  Burst 28, by contrast, shows
an extremely significant drop in frequency of approximately 0.1Hz
(0.05\%) over a timescale $\sim 1$ s.  Physical mechanisms that might
lead to such a decrease are discussed in the next section.

\section{Discussion and conclusions}
\label{disc}

Our aim in analysing the variability of the burst oscillations of
J1814 was to constrain models of the burst oscillation mechanism in
accreting millisecond pulsars.   We have focused on three aspects of
the oscillations:  fractional amplitude, 
harmonic content, and frequency.  The first two properties constrain
the size, shape and position of the emitting region, whilst 
 changes in frequency indicate motion of the emitting region.  

We can rule out the hypothesis that burst average fractional amplitude
remains constant at a level greater than 1 in $10^4$.   We can also
rule out, at a similar level, the hypothesis that the burst fractional
amplitude for all of the bursts is the same as the daily average non-burst fractional
amplitude.  The problem is in fact rather complex.  

For most of the bursts the
average fractional amplitude is in good agreement with the
accretion fractional amplitude.  Since accretion fractional amplitude
reflects the pattern of fuel deposition on the stellar surface, this
implies a low degree of fuel spread. A factor
that could restrict spreading is magnetic confinement, but the viscosity of the
accreted fuel will also be important.  If this is the case then the
hotspot size, position and inclination inferred by \citet{bha05} from
the burst oscillations should hold true for the accretion hotspot.
The geometry inferred by \citet{bha05} is that illustrated in the
center panel of Figure \ref{geometry}. 

There are however 6 bursts where the fractional amplitude is
substantially lower than the non-burst fractional amplitude, implying
a higher degree of fuel spread.  The lowest fractional amplitude burst
is the final one, which takes place in a lower accretion rate regime.
This burst shows the hallmarks of photospheric radius expansion, which
would suppress fractional amplitude.  However even if we neglect
events from the peak of the burst (where the signal is suppressed) the
fractional amplitude remains below the non-burst level.  At lower
accretion rates recurrence times should be longer, perhaps allowing a higher degree of fuel spread
between bursts.  For the remaining 5 bursts, however, it is difficult to
understand why only a subset of the bursts are affected 
if the only factor affecting spread is the time elapsed since the last
burst.  One  possibility is that the
explosive burning front of the previous burst has forced fuel out
across the stellar surface.  In this case the full burst history,
particularly the nature of the precursor burst, becomes important.  It
is however difficult to track this given the gaps in the RXTE
coverage.  

J1814 shows the strongest first harmonic content of the known
accreting millisecond pulsars, so it forms an interesting case for
analysis.  We find that the fractional
amplitude of the first harmonic is in general lower for the bursts
than for the non-burst pulsations.  We can in fact rule out
the hypothesis 
that the first harmonic content is the same as that of the non-burst emission
with a high degree of confidence (at a level greater than 1 in
$10^4$).  We cannot however rule out the
hypothesis that the fractional amplitude of the first
harmonic during the bursts remains constant over the course of the
outburst for the 27 bursts that do not exhibit photospheric radius
expansion. 

This apparent difference in behavior between the fundamental and first
harmonic is interesting. If first harmonic content is determined only
by surface emission region geometry then the discrepancy in first
harmonic content of the  burst
and non-burst emission is hard to explain, given that the
results for the fundamental suggest that the emission regions are very
similar for most of the bursts.
One possible explanation is that the first harmonic content of the
accretion-related pulsations is boosted by a contribution from
boundary layer processes.  Boundary layer emission is not apparent
during the bursts, when thermonuclear emission from the surface
dominates.  This explanation would be consistent with results for
J1808; \citet{gil98, gie02, pou03} conclude that the first harmonic
content of the non-burst pulsations of J1808 is due to Comptonized 
emission from the boundary layer.  One could test this
hypothesis by carrying out spectroscopic studies similar to those
undertaken for J1808; we leave this as a topic for future work.  It is
also interesting to note that the first harmonic content of the
non-burst emission seems to drop over the course of the outburst.  One
possible explanation for this effect is magnetic field burial caused
by accretion.  

The previous study by \citet{str03} noted apparent variations in
fractional amplitude of up to 5\% during the course of the bursts. If
genuine, such variations 
would signal rapid and rather large changes in the thermal
patterns on the 
stellar surface, something that is rather difficult to explain. On the
basis of the more detailed analysis presented in this paper we can
draw several conclusions regarding variability.  

The first is that Burst 28, the final burst, shows compelling evidence
for variation in fractional amplitude.  We can rule out the hypothesis
that the fractional amplitude remains constant at a level greater than
1 in $10^4$ for this burst. The nature of the variation is that
fractional amplitude is strongly suppressed in the peak of the
lightcurve.  This is strong evidence for photospheric radius expansion
in this burst.  

For the remaining 27 bursts, where there is no evidence for
photospheric radius expansion, we cannot rule out the hypothesis that
fractional amplitude remains constant during a burst.  Note however
that we have not tested any models that call for variation; a model
with some small ($< 1$\%) underlying variability could be a better fit
for the data.  Nonetheless this result simplifies the modelling 
requirement, as there is no need to invoke large fluctuations in
hotspot area or brightness to explain the
data.  The variations that are present appear to be random and not
dependent on progress through the burst; we detect no sign of any
consistent time-dependent effects.  

We note that variations in fractional amplitude of similar magnitude
have been observed during the bursts of several non-pulsing LMXBs
\citep{mun02}.  Our analysis shows that statistical scatter alone can
lead to apparent variations in fractional amplitude,
particularly
if one has a large sample of bursts.  The only
reliable way to assess the significance of the observed variations is
to undertake the type of analysis that we have done for J1814.  It
would be interesting to perform a similar study for the bursts of the
LMXBs, particularly since the mechanism behind the asymmetries may
differ from that operating in the pulsars. 

Some LMXBs show significant fractional amplitude decreases
in the burst rise.  The fractional amplitude shifts are thought to be the
hallmark of the spreading ignition hotspot \citep{str97, str98}.  We
find no evidence for 
fractional amplitude decrease in the burst rise, but caution that such an
effect could be masked by the accretion pulsations in this source.

The bursts of many LMXBs show evidence for frequency shifts during the
burst tail, suggesting motion of the thermal patterns on the surface.   
Our analysis confirms the earlier result of \citet{str03},
that the bursts of 
J1814 show no evidence for frequency 
shifts in the burst tail, beyond that expected due to orbital Doppler
shifts. The data are 
therefore consistent with the 
concept that the pulsed burst emission is due to the presence of a hotspot
that is near stationary in the rotating frame of the neutron star,
most likely at the visible magnetic polar cap.   

We also looked for evidence of frequency shifts in the burst rise.
None of the bursts apart from the final one show evidence of a 
statistically significant frequency shift in the burst rise.  The
final burst, however, exhibits a small ($\approx 0.1$Hz) but highly significant
frequency decrease during the burst rise.  A simple mechanism that could lead to a frequency decrease is expansion of
the burning layers.  If the layers decouple even partially from the
underlying star their 
rotation rate will drop due to conservation of angular momentum \citep{str97a}.  The
fact that the shift is so small, however, implies either a low degree
of expansion ($\sim 1$ m) or partial coupling, perhaps due to the
magnetic field.   The fact that the final burst is the most luminous
would explain why the heating and expansion is greatest in this
burst.  

The expansion model has also been suggested as an explanation for the
frequency shifts seen in some LMXB bursts (although it does have
limitations, see for example \citet{cum02}).  In a number of LMXB bursts the
frequency is observed to rise from  a low point at the peak of the
lightcurve towards 
 an asymptotic frequency in the burst tail (see
for example \citet{mun02a}).  The rise is attributed to the cooling
contracting burning layer recoupling with the underlying star and
hence spinning up.  The fact that the frequency decrease is not seen has been
attributed to the fact that the timescale on which the burning layers first
expand is very short ($\sim 10^{-6}$ s) compared to the timescale on which
photons diffuse through the atmosphere \citep{cum00}.

The timescale of the frequency shift in the rise of Burst 28 is $\sim
1$ s, far slower than the hydrostatic expansion timescale.  If
hydrostatic expansion of the burning layers is responsible, some
mechanism must be acting to slow the expansion. Although it is
conceivable that the magnetic field could act in this way, the
difference in the timescales is substantial and it seems unlikely that
the field could have such a large effect. Another possibility is that
the expansion time is slower, perhaps due to a larger amount 
of hydrogen in the burning mix compared to the LMXB bursts in which
oscillations are seen. Expansion time could also be slower if the
outer layers of the atmosphere, which expand more slowly, play a more
significant role in the emission.  A second possibility is that the
burning layers expand 
primarily laterally rather than radially, with only a limited amount
of radial expansion as the layers heat up. Lateral expansion is
plausible in a situation where the fuel is confined to an elevated
blister at the magnetic polar caps rather than being distributed 
symmetrically.  Lateral expansion could also contribute to the drop in
fractional amplitude 
towards the peak of the lightcurve.  There is however no evidence for
lateral expansion in any of the other bursts studied, as might be
expected.  The other issue associated with burning layer expansion is
that there 
needs to be some mechanism to prevent shearing in the expanding
atmosphere from washing out the asymmetry.  The shearing timescale should
be longer in a helium rich atmosphere \citep{cum00}, but the magnetic
field is an obvious candidate to reduce shear and maintain the
asymmetry in this system. 

The alternative is that we are not seeing burning layer expansion.
The observed frequency shift could instead be caused by non-radial drift of the
brightness pattern.  There are several mechanisms that could give rise to
such an effect, including the presence of a surface mode with a
non-zero pattern speed in the rotating frame (see for example
\citet{hey04}), or motion 
of a hotspot on the stellar surface (as described by
\citet{spi01}).  All of these issues need further
study. 

In addition to the issues discussed above, there are several other
areas that merit further study.  One area that we were not able to
address in much detail was the influence of the accretion and burning
history over the course of the outburst.  The level of coverage
achieved by RXTE dropped severely in the second half of the outburst,
leading to a much reduced number of burst detections. This is
unfortunate, since the only burst detected in the final portion of the
outburst (when accretion rate had dropped dramatically) seemed to
differ greatly from the other bursts.  Without more examples of bursts
in the later phase of the outburst, however, it is difficult to
determine to whether there is a major change in
behavior at late times.  If J1814 goes into outburst again it would be
highly desirable to get more even coverage throughout the outburst.       

A second question concerns the non-burst pulsations.  We have
discussed the relationship between the burst and non-burst pulsations,
pointing out both similarities and important differences.   However we
have as yet little understanding 
of the factors that control variations in the fractional amplitude and
harmonic content of the non-burst pulsations.   

We have also been unable to shed much light on the observed
variability in burst flux; in particular on the nature of the lowest
luminosity bursts.  There are hints from the data that the low
luminosity bursts may have lower average fractional amplitudes than
most of the bursts.  However, without a larger sample it is not
possible determine whether this is a statistically robust result.    

We finish on a cautionary note.  Our
analysis has shown that the data 
are consistent with there being no variability in fractional
amplitude  
over the course of a burst. However, we have not tested whether the
data are consistent with any models that call for variation, because
no quantitative models exist in the literature at the present time. A
similar analysis 
could easily be performed should such a model be developed.  Our study
also illustrates the need for high PCU coverage.  J1814 is a
relatively faint burst source, and we are trying to measure
short-timescale variations in quantities that vary widely due to
statistical effects alone. Most of the bursts in the 2003 outburst
were observed with only 3 PCUs.  Increasing the number of PCUs for
future observations of this source would allow far more stringent
tests of the hypotheses examined in this paper.

\acknowledgments
This research has made use of data obtained from the High
Energy Astrophysics Science Archive Research Center (HEASARC) provided
by NASA's Goddard Space Flight Center.  We would also like to thank
the anonymous referee for a careful reading of the paper and helpful
suggestions.

\clearpage

\LongTables 
\begin{landscape}

  \begin{deluxetable}{ccccccccccccc}
  \tabletypesize{\scriptsize}
  \tablecaption{Summary of burst properties. \label{bdatatab}}
  \tablewidth{0pt}
  \tablehead{\colhead{Index} & \colhead{Peak time}  &
  \colhead{Peak flux} &
  \colhead{$\nu$} & \colhead{Rise time} & \colhead{}  &
  \multicolumn{2}{c}{RMS fractional 
  amplitude (\%)} & \colhead{}  & \multicolumn{2}{c}{Variability} &
  \colhead{}& \colhead{$N_\mathrm{acc}$/$N_\mathrm{bur}$} \\
\cline{7-8} \cline{10-11}\\ \colhead{}  & \colhead{(Days - MJD 52790)}
  & \colhead{($10^2$
  Cts $\mathrm{s}^{-1} \mathrm{PCU}^{-1}$)} &
  \colhead{(Hz)} & \colhead{(s)} &  \colhead{}  &
  \colhead{Fundamental} & 
  \colhead{1st harmonic} & \colhead{}  & \colhead{$V_r$}  &
  \colhead{$V_f$} & \colhead{} & \colhead{}  
  }
  \startdata

  1 & 6.0407 & 3.8 & 314.314 & 7.14 & & $7.8 \pm 0.5$ & $1.7^{+ 0.6}_{- 0.5}$
 & &  0.297 & 0.191 & & 0.18 \\[4pt]

  2 & 7.2687 & 9.9 & 314.307 & 3.00 & & $10.1 \pm 0.4$ & $2.9^{+0.4}_{-0.4}$
  & & 0.768 & 0.421 & & 0.09 \\[4pt]

  3 & 7.8836 & 11.1 & 314.404 & 2.91 & &  $10.1 \pm 0.4$ & $2.3^{+0.4}_{-0.3}$
  & &  0.792 & 0.471 & & 0.09 \\[4pt] 

  4 & 9.1963 & 10.7 & 314.338 & 3.22 & & $9.7 \pm 0.3$ & $2.4^{+0.4}_{-0.3}$  &&  0.435 & 0.910 & & 0.09 \\[4pt]

  5 & 10.0993 & 10.7 & 314.319 & 3.32 & & $11.0 \pm 0.5$ & $2.0^{+0.5}_{-0.4}$&&  0.448 & 0.605 & & 0.11  \\[4pt]

 6 & 11.0293 & 3.6 & 314.316 & 5.84 & & $8.7 \pm 0.6$ & $2.0^{+0.7}_{-0.5}$ &&  0.319 & 0.097 & & 0.20 \\ [4pt]

  7 & 12.4664 & 7.1 & 314.335 & 4.17 & & $11.4 \pm 0.3$ & $2.3^{+0.3}_{-0.3}$ &&  0.483 & 0.134  & & 0.14  \\ [4pt]

  8 & 12.5633 & 4.0 & 314.365 & 5.17 & & $10.8 \pm 0.6$  &
  $2.3^{+0.6}_{-0.5}$ &&  0.833 & 0.114 & & 0.19  \\ [4pt]

  9 & 13.0594 & 3.7 & 314.406 & 5.16 & & $8.8 \pm 0.5$ & $2.4^{+0.5}_{-0.4}$ & &  0.699 & 0.549 & & 0.18 \\ [4pt]

  10 & 13.7342 & 7.9 & 314.370 & 4.51 & & $11.4 \pm 0.3$ &
  $1.7^{+0.4}_{-0.3}$ &&  0.025 & 0.809  & & 0.13  \\ [4pt]

  11 & 14.0145 & 8.7 & 314.324 & 3.25 & & $11.0 \pm 0.3$ & $2.0^{+0.4}_{-0.3}$ &&  0.485 & 0.824  & & 0.13 \\ [4pt]
 
  12 & 15.7895 & 6.6 & 314.331 & 4.37 & & $11.0 \pm 0.3$ &
  $1.6^{+0.4}_{-0.3}$ &&  0.168 & 0.325 & &  0.13 \\[4pt] 

  13 & 16.7475 & 6.8 & 314.346 & 4.70 & & $12.2 \pm 0.4$ & $2.5^{+0.4}_{-0.3}$ &&  0.118 & 0.968 & & 0.15  \\[4pt]

  14 & 16.8181 & 3.5 & 314.396 & 8.01 & & $10.7 \pm 0.8$ & $0.8^{+1.4}_{-0.8}$ &&  0.693  & 0.514 & & 0.22 \\[4pt]

  15 & 17.6670 & 7.1 & 314.393 & 4.42 & & $11.2 \pm 0.3$ & $2.1^{+0.3}_{-0.3}$ & &  0.368 & 0.295 & & 0.13 \\ [4pt]

  16 & 18.7955 & 7.3 & 314.368 & 4.35 & &  $10.8 \pm 0.4$ & $1.5^{+0.4}_{-0.3}$ & &  0.275 & 0.608 & & 0.12 \\[4pt]

  17 & 19.7817 & 7.6 & 314.356 & 4.54 & & $10.8 \pm 0.4$ & $2.7^{+0.4}_{-0.3}$  &&  0.217 & 0.352  & & 0.12 \\[4pt]

  18 & 20.0682 & 10.2 & 314.327 & 3.26 & &  $10.6 \pm 0.3$ & $2.0^{+0.4}_{-0.3}$ & &  0.073 & 0.918 & & 0.11 \\[4pt]

  19 & 20.9015 & 3.1 & 314.407 & 6.98 & &  $ 10.6 \pm 0.6$ & $2.1^{+0.7}_{-0.5}$ & &  0.668 & 0.419 & & 0.24 \\ [4pt]

  20 & 21.6396 & 9.8 & 314.378 & 3.99 & & $10.2 \pm 0.4$ & $1.9^{+0.4}_{-0.3}$ &&  0.171 & 0.881 & & 0.12 \\[4pt]

  21 & 22.8887 & 8.5 & 314.375 & 4.74 & &  $11.0 \pm 0.4$ & $2.2^{+0.4}_{-0.3}$ && 0.512 & 0.777 & & 0.11 \\[4pt]

  22 & 23.4688 & 6.7 & 314.308 & 4.74 & & $11.3 \pm 0.3$ & $2.1^{+0.3}_{-0.3}$& &  0.398 & 0.968 & & 0.14 \\[4pt]

  23 & 27.6937 & 6.9 & 314.379 & 4.41 & &  $11.1 \pm 0.3$ & $1.4^{+0.4}_{-0.3}$ &&  0.038 & 0.370 & & 0.11 \\[4pt]

  24 & 27.8834 & 7.2 & 314.360 & 5.08 & & $11.4 \pm 0.4$ &
  $2.0^{+0.4}_{-0.3}$ & &  0.852 & 0.678 & &  0.12 \\[4pt]

  25 & 28.8537 & 6.8 & 314.338 & 6.09 & & $ 11.1 \pm 0.5$ & $1.6^{+0.5}_{-0.4}$ & &  0.372 & 0.259  & & 0.13 \\[4pt]

  26 & 37.2535 & 7.0 & 314.383 & 4.33 & &  $11.1 \pm 0.3$ & $1.4^{+0.3}_{-0.3}$&&  0.141 & 0.856  & &  0.13  \\[4pt]

  27 & 38.7935 & 6.9 & 314.305 & 4.14 & &  $10.9 \pm 0.4$ & $1.6^{+0.4}_{-0.3}$ & &  0.698 & 0.508  & & 0.12 \\[4pt]

  28 & 47.7381 & 16.0 & 314.353 & 2.53 & & $3.5 \pm 0.3$ & $0.0^{+0.5}$ &  &  $<10^{-4}$ & 0.030 & & 0.03 \\
  \enddata
  \end{deluxetable}
\clearpage
\end{landscape}

\end{document}